\documentclass[fleqn,usenatbib]{mnras}

\usepackage{newtxtext,newtxmath}

\usepackage[T1]{fontenc}

\DeclareRobustCommand{\VAN}[3]{#2}
\let\VANthebibliography\thebibliography
\def\thebibliography{\DeclareRobustCommand{\VAN}[3]{##3}\VANthebibliography}

\usepackage{graphicx}	
\usepackage{amsmath}	
\usepackage{multicol, blindtext, multirow}
\usepackage{xcolor}
\usepackage{siunitx}
\usepackage{gensymb}

\title[Precessing Radio Jets: S-shaped morphology]{Dynamical Modelling and Emission Signatures of a Candidate Dual AGN with Precessing Radio Jets}

\author[Giri et al. 2022]{
Gourab Giri,$^{1}$\thanks{E-mail: gourab@iiti.ac.in}
Ravi Pratap Dubey,$^{1,2}$
K. Rubinur,$^{3}$
Bhargav Vaidya,$^{1}$\thanks{E-mail: bvaidya@iiti.ac.in}
Preeti Kharb$^{3}$
\\
$^{1}$Department of Astronomy, Astrophysics and Space Engineering, Indian Institute of Technology Indore, Simrol 453552, India\\
$^{2}$Max Planck Institute for Astronomy, K\"onigstuhl 17, D-69117 Heidelberg, Germany\\
$^{3}$National Centre for Radio Astrophysics (NCRA-TIFR), Pune University Campus, Ganeshkhind, Pune 411007, India 
}

\date{Accepted XXX. Received YYY; in original form ZZZ}

\pubyear{2021}
\defcitealias{rubinur2017}{R17}
\defcitealias{hjellming1981}{HJ81}
\newcommand{\fpath}{figures/}
\begin{document}
\label{firstpage}
\pagerange{\pageref{firstpage}--\pageref{lastpage}}
\maketitle

\begin{abstract}
In this paper, we have modelled the dynamical and emission properties (in the presence of radiative losses and diffusive shock acceleration) of an observed S-shaped radio source (2MASX J12032061+131931) due to a precessing jet. In this regard, we have performed high-resolution 3D magnetohydrodynamic (MHD) simulations of a precessing jet in a galactic environment. We show the appearance of a distinct S-shape with two bright hotspots when the bow shock region weakens over time. The formed morphology is sensitive to the parameter selections. The increased interaction between the helical jet and the ambient medium and the deceleration of the jet due to MHD instabilities also greatly affect the resulting structure. Hence, kinematic models must be corrected for these deceleration effects in order to adequately predict the precession parameters. The synthetic spectral index map shows that the jet side and leading edges possess relatively steeper spectral index values than the jet ridge lines, whereas the hotspots show flat spectral index values. The jets are also found to be highly linearly polarized (up to 76\%) and the magnetic field lines, in general, follow the jet locus which is formed due to the jet-ambient medium interaction. Diffusive shocks, in this context, keep the structure active during its course of evolution. Furthermore, we have demonstrated that these galaxies deviate significantly from the `equipartition' approximation leading to a discrepancy in their spectral and dynamical age.

\end{abstract}

\begin{keywords}
galaxies: jets -- galaxies: individual: 2MASXJ12032061+1319316 -- radiation mechanisms: non-thermal -- (magnetohydrodynamics) MHD -- methods: numerical
\end{keywords}



\section{Introduction}
Active galactic nuclei (AGN) are the centre of galaxies with  accreting supermassive black holes (SMBHs). 
A small fraction of them show the presence of radio jets ranging from sub-kiloparsec (kpc) to megaparsec (Mpc) scales \citep{kellermann1989,Bassani2016,Dabhade2020}. 
Some of these radio galaxies show systematic change in their jet propagation direction, forming tailed or winged sources \citep{Hardcastle2019,Cotton2020,Muller2021,Pandge2021}. It is now believed that the underlying origin of tailed radio galaxies is the motion of the host galaxy with respect to the ambient medium \citep{Oneill2019}. This produces a mirror symmetry of the jets by bending them in the same direction. On the other hand, the winged morphology shows a peculiar inversion symmetry as the jets bend in opposite directions, forming X-, S- or Z-shaped structures \citep[see][]{ekers1982}.

S-shaped radio galaxies are a sub-class of winged sources that can be identified by observing the presence of a distinctly curved jet \citep{Gower1982,Krause2019}. However, it is often difficult to isolate S-shaped sources from winged sources having Z- or X-shaped morphologies by only identifying a misaligned jet from the lobe axis. For example, the observed bent jet structure of NGC 326 has been attributed to S-, Z- and X-shaped structures in the studies by \citet{Ekers1978,Gopal-Krishna2003,Hodges-kluck2012,Hardcastle2019}. As a result of this ambiguity in their classification, formation process of such galaxies is still under debate. In particular, for X-shaped radio galaxies, the off-axis bent jet morphology can be explained by the back-flow mechanism (back-flowing jet plasma deflected laterally by the asymmetric ambient medium) \citep{Leahy1984,Capetti2002,Rossi2017,Giri2022} or by the sudden spin-flip mechanism (sudden spin-flip due to BH-BH coalescence) \citep{Dennett-Thorpe2002,Rottmann2002,Hodges-kluck2010}. On the other hand, the presence of a precessing jet is often alluded to describe S- or Z-shaped morphologies \citep{Rottmann2002}.

The plausible sources that could influence the curved jet structure include: a) presence of a binary AGN (separation < 100 pc) or b) existence of a dual AGN (separation 0.1-10 kpc) \citep{begelman1980,rubinur2017,rubinur2021} or c) simply an unstable accretion disk \citep{Pringle1996,Kurosawa2008,Liska2018}. 
Galaxies having SMBH at their centres during mergers, or a minor merger resulting in a massive gas inflow to the central AGN can trigger these aforementioned activities leading to jet precession \citep{Rottmann2002}.
However, the direct evidence for observing these mechanisms is somewhat elusive as the scales are typically inaccessible to present-day telescopes \citep{rubinur2017,Krause2019,Kharb2019}. In some peculiar cases, the coexistence of dual AGN and S-shaped jet has been observed. The classic example is that of NGC 326. This radio galaxy has a confirmed dual AGN \citep{Murgia2001}, and it shows the presence of a curved jet that has recently been studied by \citet{Hardcastle2019} who demonstrated the vital role of the ambient medium in shaping the jet structure. In general, identifying the dual AGN candidates via observation is a challenging task as they may have degenerate signatures. For example, the double-peaked AGN (DPAGN) emission lines observed in the optical nuclear spectra, which are thought to be a potential observational signature of dual AGN \citep[DAGN;][]{Fu2011}, have been found to have different origins like jet-medium interaction or a rotating gaseous disk \citep{Fu2012,Kharb2015,rubinur2019,Kharb2019}.
The above observational studies of winged radio sources have depicted the complex nature of S-shaped jets whose formation mechanism is still unclear. With this motivation, we aim to numerically model a particular dual AGN candidate and comprehend if the curved jet observed can be formed due to underlying precession. 

One of the early numerical studies to understand the dynamical signatures from precession jets based on several precession parameters has been carried out by
\citet[][]{Hardee2001}. In recent times, numerical models of jet precession have been used to adequately compare with observed sources. 
For example, \citet{Nawaz2016} have numerically reproduced the morphology observed in Hydra A. Numerical simulations of precessing jets by \citet{Falceta2010} have been used to obtain the X-ray cavities observed in the Perseus galaxy cluster. 
The study led by \citet{horton2020} showed the formation of multiple S-shaped sources based on jet precession that has been reported in the observational study of \citet{Krause2019}. Furthermore, detailed numerical studies were also performed to better understand the precessional motion of an X-ray binary source SS~433 \citep{Monceau2014,Monceau2015}. 
 
In-spite of detailed modelling of the dynamical aspects of precessing jets, comparison with observed sources primarily relies on using proxies (e.g. pressure as emissivity). 
Recently, several studies have incorporated non-thermal particles in an Eulerian-Lagrangian hybrid framework for providing an alternative and improved method of comparing dynamical simulations with observations \citep{Oneill2019,Borse2021,Mukherjee2021,Giri2022,Yates2022}. Such hybrid modelling evolves the particle spectra in the presence of radiative and adiabatic losses or the particle re-energization processes, which continuously influence the distribution of relativistic electrons (hence the emission) as they move in the cocoon. 
The goal of the present work is to verify if a precessing jet can produce the observed S-morphology of a particular Seyfert galaxy using the Eulerian-Lagrangian hybrid framework that incorporates effects due to radiative and adiabatic losses in the presence of diffusive shock acceleration.

\citet{rubinur2017} (\citetalias{rubinur2017} hereafter) have detected the symmetric S-shaped radio jets of the double-peaked emission galaxy 2MASX J12032061+1319316 (2MASX J1203 hereafter) from their multi-wavelength radio observations with the EVLA \footnote{Expanded VLA: \url{http://www.aoc.nrao.edu/evla/}}. The 6 and 15 GHz images showed two lobes along with hotspots on each side of the nucleus, while 8.5 and 11.5 GHz images revealed a core situated in between the lobes. \citetalias{rubinur2017} fitted a jet precession model proposed by \citet{hjellming1981} \citepalias[hereafter][]{hjellming1981} on 2MASX J1203 and found a precession timescale of $\sim10^{5}$ yrs. This matched with the jet lifetime calculated from spectral ageing using radio flux density and spectral index values. 
We specifically focus on reproducing the observed morphology using 3D magneto-hydrodynamic (MHD) simulations. The objective would be to  provide support for or dismiss the precessing jet model as a formation mechanism of this particular S-shaped source. 
We also intend to explore the parametric space under which the formation of this S-shaped source is feasible and also provide a synthetic polarisation map that can serve as a template for further observational studies of this source.

This paper is arranged as follows: in Section \ref{Numerical Setup}, we describe our numerical setup considering both the dynamical as well as the emission perspective of it. The dynamical evolution and characteristics of different simulation runs are discussed in Section \ref{sec:dynamics}. Associated spectral properties of the formed structures are highlighted in Section \ref{Spectral signatures} that includes intensity measures, spectral behaviour study, and polarization map study of the radio structure. We further investigate the galaxy's equipartition state and its impact on age estimation in Section \ref{Equipartition approximation and Age estimation}. Finally, we highlight the key conclusions drawn from our work in Section \ref{sec:summary}.

\section{Numerical Setup} \label{Numerical Setup}

To perform our numerical simulations, we have used the PLUTO code \citep{mignone2007} which solves and updates with time the set of magneto-hydrodynamic equations defined as
 \begin{align}
     \frac{\partial \rho}{\partial t} + \textbf{v} \cdot \nabla \rho + \rho \nabla \cdot \textbf{v} = 0 \\
     \frac{\partial \textbf{v}}{\partial t} + \textbf{v} \cdot \nabla \textbf{v} + \frac{1}{\rho}\textbf{B} \times(\nabla \times \textbf{B}) + \frac{1}{\rho}\nabla P = 0 \\
     \frac{\partial \textbf{B}}{\partial t} + \textbf{B}(\nabla \cdot \textbf{v})-(\textbf{B} \cdot \nabla)\textbf{v}+(\textbf{v}\cdot \nabla)\textbf{B} = 0 \\
     \frac{\partial P}{\partial t} + \textbf{v} \cdot \nabla P+ \rho c_s^2 \nabla \cdot \textbf{v} = 0 
 \end{align}
Here $\rho$, \textbf{v}, \textbf{B} and $P$ represent the mass density, velocity, magnetic field and thermal pressure of the system respectively.
These equations are solved with second-order accuracy in space using linear reconstruction and HLLC Riemann solver \citep{Mignone2006}. We have carried out our full-scale 3D simulations in the Cartesian geometry and have adopted an ideal equation of state where the adiabatic constant $\Gamma = 5/3$. We ensured the solenoidal condition of the magnetic field (i.e. $\nabla \cdot {\rm \textbf{B}} = 0$) by adopting the method of divergence cleaning as suggested by \citet{Dedner2002}.

\subsection{Dynamical model}
\label{sec:dynsetup}
The ambient condition of our numerical setup is modelled using the spherical King's density profile \citep{cavaliere1978} defined as
\begin{equation}
    \label{eq:kings}
    \rho (r) = \rho_0 \left(1+\left(\frac{r}{r_{c}}\right)^2\right)^{-3/2}
\end{equation}
where $\rho_{0}$ is the density of the galaxy at $r= 0$ and adopted to be $10 \times (1.66\times 10^{-24})$ gm/cc i.e., 10 particles/cc \citep[][considering only gas mass]{Querejeta2015}. Using this density profile, we are modelling the bulge of an S0 galaxy as the ambient morphology of 2MASX J1203 is suggested to be an S0 type \citepalias{rubinur2017}. Here, $r$ ($= \sqrt{x^2 +y^2+z^2}$) represents the radial distance from the centre of the galaxy and $r_c$ represents the core radius taken to be 0.7 kpc \citep{Querejeta2015}. The associated pressure profile ($P_a$) is kept to be constant all over the domain so that a static equilibrium is obtained for the ambient medium (initially) in the absence of gravity. The value of $P_a$ is $2.7 \times 10^{-10}$ $\rm dyn/cm^2$, chosen such that the injected jet remains in the thermal pressure balance with the ambient medium initially. As the density varies in the ambient medium, the temperature in the medium shows radial variation such that its average value is obtained as $0.6\times10^6$ K. The sound speed corresponding to the average ambient temperature is estimated to be $1.26\times10^7$ cm/s. 
The simulation domain extends from -1 kpc to 1 kpc in the $x$- and $z$-direction and from -2 kpc to 2 kpc in the $y$-direction distributed uniformly in  $ 128 \times 256 \times 128$ grids (along the $x \times y \times z$ direction). We put an outflow boundary condition at all boundaries of the domain.

Further, we have introduced a cylindrical jet nozzle at the origin (i.e. from the centre of the ambient medium) which continuously injects an underdense jet having velocity ${\rm v}_j$. The jet is underdense by a factor of $0.01$ with respect to the ambient medium i.e. $\rho_j = 0.01\rho_0$ \citep{Kharb2019}.
The cylindrical injection region has a size of $200\ {\rm pc} \times 160\ {\rm pc}$ (radius $\times$ length) which is further divided into two parts to incorporate a bi-directional jet into the domain. The counter-jet region (from the origin towards the -ve $y$-direction) and the jet-region (from the origin towards the +ve $y$-direction) inject the same underdense jet with its velocity reversed. With the chosen resolution, we ensured 26 grid cells within the jet diameter that can resolve the transverse structures properly. We note here that we have optimized our numerical simulations and hence the results noted in this work by conducting a convergence study in terms of resolution of our runs (discussed further in Section \ref{sec:High Resolution}).

\begin{figure}
    \centering
    \includegraphics[width=\columnwidth]{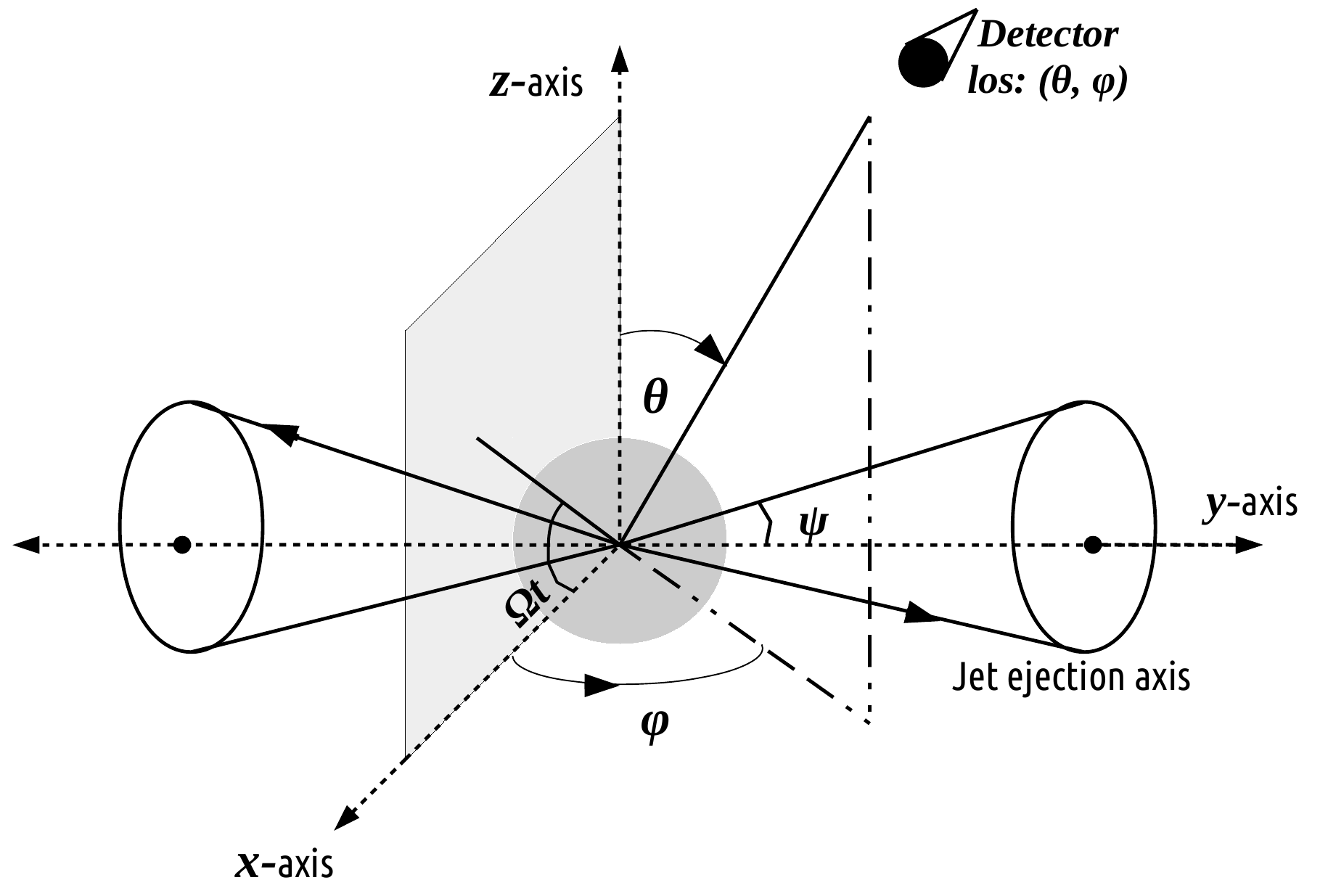}
    \caption{A 3D representative image showing the jet-ambient medium configuration used in our study. The bi-directional jet undergoes a precessional motion around the $y$-axis at an angle $\psi$ with a precession period $\tau$, resulting in an angular velocity $\Omega=2\pi / \tau$ in the $x-z$ plane. The line of sight visualization angle ($\theta, \upvarphi$) is also overlaid here, along which the synchrotron emissivities were calculated. The central sphere is a representative bulge of an S0 galaxy, inside of which the jet performs its complex motion.} 
    \label{fig:cartoon}
\end{figure}

To imitate the precession of the jet, we have followed the geometrical model proposed by \citetalias{hjellming1981} which is also used in modelling the precessing jet of 2MASX J1203 \citepalias{rubinur2017}.
Our jet undergoes a precessional motion around the $y$-axis at an angle $\psi$ with a precession period $\tau$, resulting in an angular velocity $\Omega=2\pi / \tau$ in the $x-z$ plane. 
Based on this configuration, the velocity components of the jet along the three Cartesian axes can be written as 
\begin{equation}
({\rm{v}}_x, \ {\rm{v}}_y, \ {\rm{v}}_z) \ = \ ({\rm{v}}_j\sin{\psi}\cos{\Omega t}, \ {\rm{v}}_j\cos{\psi}, \ {\rm{v}}_j\sin{\psi}\sin{\Omega t})
\end{equation}
where $t$ represents the time of its evolution. The configuration is highlighted in Fig. \ref{fig:cartoon} via a representative diagram. 
In addition to this, we also have incorporated a toroidal magnetic field ($B_{\phi}$) in the jet having a form similar to that mentioned in \citet{Lind1989} i.e.
\begin{equation} \label{eq1}
\begin{split}
B_{\phi} & = B_{i} \frac{\mathcal{R}}{a} \text{ \space \space \space \space \space \space  for $\mathcal{R}$ $<$ $a$} \\
 & = B_{i} \frac{a}{\mathcal{R}} \text{ \space \space \space  \space \space \space  for $\mathcal{R}$ $\geq$ $a$} 
\end{split}
\end{equation}
Here $B_i$ is constant, $a$ is the jet magnetization radius chosen to be 0.8 times the jet radius and $\mathcal{R}=\sqrt{x^2+z^2}$, $\phi = \tan^{-1}(z/x)$ are the usual cylindrical coordinates obtained in the $x-z$ plane. We have chosen the value of $B_i$ and $a$ such that we get an average magnetic field of 0.19 mG in the jet base, which is typical for AGN outflows \citep{Fan2008}. However, we have also varied this value slightly higher (1.6 times) and slightly lower (2 times) in order to observe its effect on the formed morphology. We name these runs as \textit{CaseB\_hB} and \textit{CaseB\_lB} respectively.
In the counter-jet region, the direction of the magnetic field is opposite to that in the jet region.
To maintain the radial equilibrium in the jet, we obtained the pressure distribution in the jet ($P_{j}$) by solving the radial momentum balance equation between thermal, centrifugal and magnetic forces  \citep{Lind1989,tesileanu2008} as
\begin{equation}
     P_{j}(\mathcal{R}) = P_{a} + \frac{\rho_j\ \rm{v}_{\phi}^2}{2} + B_i^2 \left(1-{\rm{min}}\left(1,\frac{\mathcal{R}^2}{a^2}\right) \right)
\end{equation} 
where ${\rm{v}}_{\phi} = \mathcal{R} \Omega$. 

We note here that the injected jet nozzle in our simulation does not have an opening angle, as we expect the S-shaped morphology to be governed primarily by the jet dynamics, which depend mainly on the precession parameters.

Based on this configuration, we have run our first setup by adopting parameters from \citetalias{rubinur2017} that provides ${\rm v}_j$ as 0.023 times light speed ($c$),  $\psi$ as $21^{\circ}$ and $\tau$ as $0.95 \times 10^5$ yrs. This run is labelled as \textit{CaseA} for further references. As we also wanted to explore the parameter space under which formation of these winged galaxies is feasible, we have conducted another run using parameters ${\rm v}_j = 1.25 \times 0.023c$, $\psi =$ $25^{\circ}$ and $\tau =$ $0.95 \times 10^5$ yrs (simulation label: \textit{CaseB}). In addition to this, two runs under \textit{CaseB} were performed where the precession period is varied 1.5 times higher (\textit{CaseB\_hPP}) and 1.5 times lower (\textit{CaseB\_lPP}) than $0.95 \times 10^5$ yrs (see Table \ref{tab:simsetups}). These additional runs have the potential to infer about the central engine of the galaxy. One point to note here is that all of our simulations were conducted in dimensionless units, where the units for their conversion back into physical scales are chosen by keeping consistency with the observational data. We choose units for three basic parameters i.e. length as 200 pc (jet radius), velocity as $5.25 \times 10^7$ cm/s (sound speed in the jet assuming jet temperature as $10^7$ K) and density as $10^{-3}$ \si{amu/cc}, based on which the units for all other parameters can be defined. For example, the unit for time can be obtained as (Unit length/Unit velocity) which is 0.37 Myr. All of our simulations ran for a time of 2.98 Myr, and the data were saved in each 0.19 Myr interval.

The kinetic power ($L_{\rm kin}$) of our injected jet is evaluated using the equation below (in simulation units)
\begin{equation}
    L_{\rm kin} = \pi \mathcal{R}_j^2 \mathcal{M}_j \bigg(\frac{1}{2}\mathcal{M}_j^2 \rho_j + \frac{5}{2} T_j \rho_j + \frac{B^2}{2}\bigg)
\end{equation}
where $\mathcal{R}_j$, $\mathcal{M}_j$, $\rho_j$, $T_j$ and $B$ are the jet radius, jet mach number, jet density, jet temperature and magnetic field in the jet base respectively \citep{hardcastle2014}.
Based on the choice of parameters mentioned above, the value of $L_{\rm kin}$ is estimated to be $3.3 \times 10^{43}$ erg/s for \textit{CaseA} and $6.6 \times 10^{43}$ erg/s for \textit{CaseB}. Both of these numbers are close to the value estimated for the source 2MASX J1203 i.e. $4.9 \times 10^{43}$ erg/s, using the empirical relation as shown by \citet{Merloni2007}. We note that the injected jet is kinetically dominated (by two orders of magnitude) compared to thermal and magnetic energy, which are roughly comparable. 

\subsection{Emission model} \label{Emission model}
We make use of the hybrid framework of PLUTO code \citep{Vaidya_2018,Mukherjee2021} in order to inject the Lagrangian macro particles into the domain to model the non-thermal emission of the galaxy. We have injected these particles through the jet nozzle which further follows the fluid streamlines. We have chosen an injection rate such that we obtained nearly $10^6$ particles at the end of the simulation that can model the emission features adequately. Each Lagrangian particle is an ensemble of non-thermal electrons distributed according to a power-law defined as $N(\gamma) = N_0 \gamma^{-p}$ under the limiting Lorentz factor $\gamma_{\rm min}$ ($10^2$) to $\gamma_{\rm max}$ ($10^{10}$). Here $p$ is the power-law index which will be updated for the particles depending on the micro-physical processes they encounter. The initial choice of $p$ is set to the values of 3 or 4 for different simulation runs as has been highlighted in Table \ref{tab:simsetups}. The value of $N_0$ can be obtained from $\int_{\gamma_{\rm min}}^{\gamma_{\rm max}}N_0 \gamma^{-p} d\gamma = n_{\rm micro}$ where $n_{\rm micro}$ is the non-thermal electron number density. To quantify $n_{\rm micro}$, we first assumed that the galaxy is in an equipartition stage, where the magnetic energy density in the jet base $\Big(\frac{B_{\rm dyn}^2}{8\pi}  \Big)$ is in an equal share with the radiating electron's energy density ($U_e$). The equipartition condition in cgs units can be written as \citep{Hardcastle2002,Hardcastle2010}
\begin{equation}
\label{eq:equ}
U_e = m_e c^2 \ \int_{\gamma_{\rm min}}^{\gamma_{\rm max}} \gamma N(\gamma) d\gamma \left(= \frac{B_{\rm eq}^2}{8\pi}\right) = \epsilon \frac{B_{\rm dyn}^2}{8\pi}
\end{equation}
where $m_e c^2$ represents the rest mass energy of the electron, $B_{\rm eq}$ is the equipartition magnetic field strength and $\epsilon = 1$ for the equipartition state (Sim. label: \textit{CaseB\_eqp}). We have also adopted a value of $\epsilon = 1/30$ which corresponds to an under-equipartition state of the galaxy (see Table \ref{tab:simsetups}). It is important to highlight here that $B_{\rm eq}$ is merely a proxy for the magnetic field defined in such a way that the magnetic energy associated with it is equal to the energy of the radiating electrons and so cannot be considered as the physical magnetic field. We denote the terms under-equipartition if $B_{\rm eq} < B_{\rm dyn}$, equipartition if $B_{\rm eq} = B_{\rm dyn}$ and above-equipartition if $B_{\rm eq} > B_{\rm dyn}$. Using the initial distribution of $N(\gamma)$, $B_{\rm dyn}$ and $\epsilon$, we have estimated the values of $n_{\rm micro}$ for each cases which are highlighted in Table \ref{tab:simsetups}.

This initial condition will subsequently get updated for the particles as they undergo various micro-physical processes considered here i.e. the adiabatic and radiative losses and the diffusive shock acceleration. In this regard, one should note that the first shock that most of the particles encounter occurs close to the injection region due to the jet-ambient medium interaction and recollimation. The radiative losses considered here are the synchrotron and the Inverse-Compton (due to cosmic microwave background photons) losses for which we have adopted the redshift of the galaxy as, $z = 0.0584$ \citepalias{rubinur2017}. We have carried out our emission modelling in four synchrotron frequencies ($\nu$) i.e. in 6 GHz, 8.5 GHz, 11.5 GHz, and 15 GHz. The synchrotron emissivities at these frequencies are evaluated along the line of sight angle of ($\theta,\ \upvarphi$) which is demonstrated via the representative diagram in Fig. \ref{fig:cartoon}. Intensity ($I_{\nu}$) maps are then obtained by integrating the emissivities along that viewing direction and rotating the image by an angle $\chi$ for visualization. For \textit{CaseA} the values of ($\theta,\ \upvarphi, \ \chi$) are ($90^{\circ},\ 38^{\circ},\ 33^{\circ}$) \citepalias[as obtained for our configuration from][]{rubinur2017}, whereas the values of ($\theta,\ \upvarphi, \ \chi$) for rest of the runs are set as ($90^{\circ},\ 20^{\circ}, 19^{\circ}$). The hybrid PLUTO code also provides Stokes parameters (i.e., $Q_{\nu}$ and $U_{\nu}$) as output from linearly polarized synchrotron radiation. Using these parameters, we have studied the polarization state of the galaxy at $\nu = 11.5$ GHz.

\begin{table*}
    \centering
    \caption{Here, we have listed parameters that have been varied in our simulations, raising a total of 8 runs. The 1st column represents the simulation labels attached to each run. The 2nd, 3rd, 4th, 5th, and 6th columns represent the jet speed (${\rm v}_j$), corresponding Mach number ($\mathcal{M}_j$), the tilt angle ($\psi$) of the jet, the precession period ($\tau$), and average magnetic field in the jet base ($B_{\rm av}$), respectively. The 7th, 8th, 9th, and 10th columns represent the initial power-law index of electrons ($p$), deviation factor from equipartition ($\epsilon$), the non-thermal particle density in the jet ($n_{\rm micro}$), and jet power ($L_{\rm kin}$), respectively. At the end (column 11th), we have highlighted the properties of each run that has been targeted in that simulation, justifying their names in column 1. We run all of our simulations till 2.98 Myr, and the data are saved in each 0.19 Myr interval.}
    \begin{tabular}{|r|c|c|c|c|c|c|c|c|c|c|}
         \hline
         \hline
         Sim. label & Jet speed & Mach  & Tilt angle& $\tau$ &$B_{\rm av}$ & $p$ & $\epsilon$ & $n_{\rm{micro}}$ & Jet Power&Remarks  \\
          & (${\rm v}_j$) &number ($\mathcal{M}_j$)& $\psi$ ($^{\circ}$)  & (Myr)& (mG) & &  & (part./cc)&$L_{\rm kin}$ (erg/s)& \\
         \hline
         \textit{CaseA} & $0.023c$ &13.14& 21 & 0.095 &0.19& 3 & 1/30 & $3.3 \times 10^{-7}$&$3.3 \times 10^{43}$&Parameters from R17  \\
         \textit{CaseB} & $0.0288c$ &16.43& 25 & 0.095 & 0.19 & 3 & 1/30 & $3.0 \times 10^{-7}$& $6.6 \times 10^{43}$&Modified run \\
         \textit{CaseB\_hPP} & $0.0288c$ &16.43& 25 & 0.143&0.19 & 3 & 1/30 & $3.0 \times 10^{-7}$& $6.6 \times 10^{43}$&High precession period   \\
         \textit{CaseB\_lPP} & $0.0288c$ &16.43& 25 & 0.063&0.19 & 3 & 1/30 & $3.0 \times 10^{-7}$& $6.6 \times 10^{43}$& Low precession period \\
         \textit{CaseB\_eqp} & $0.0288c$ &16.43& 25 & 0.095&0.19 & 3 & 1 & $8.9\times10^{-6}$& $6.6 \times 10^{43}$& Equipartition assumed\\
         \textit{CaseB\_p4} & $0.0288c$ &16.43& 25 & 0.095&0.19 & 4 & 1/30 & $3.9\times10^{-7}$& $6.6 \times 10^{43}$&Steeper particle spectra \\
         \textit{CaseB\_hB} & $0.0288c$ &16.43& 25 & 0.095 & 0.3 & 3 & 1/30 & $8.5 \times 10^{-7}$& $6.6 \times 10^{43}$&High B-field  \\
         \textit{CaseB\_lB} & $0.0288c$ &16.43& 25 & 0.095 & 0.095 & 3 & 1/30 & $8.3 \times 10^{-8}$& $6.6 \times 10^{43}$& Low B-field  \\
         \hline
    \end{tabular}
    \label{tab:simsetups}
\end{table*}

\section{Dynamical evolution: S-shape formation} \label{sec:dynamics}

\begin{figure*}
    \centering
    \includegraphics[width=2\columnwidth,height=6cm]{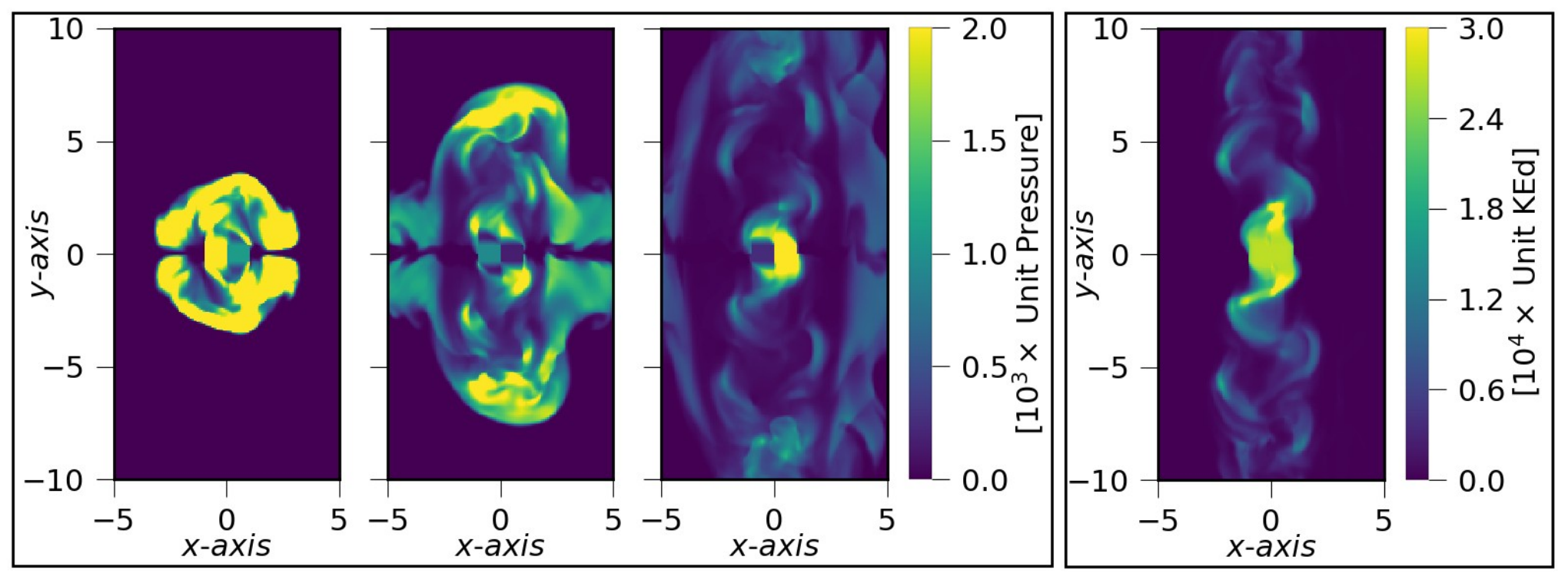}
    \caption{\textit{Left panel:} time evolution of pressure distribution $P (x, y;\ z = 0)$ in the cocoon at time 0.75 \si{Myr} (\textit{left}), 1.5 \si{Myr} (\textit{centre}) and 2.23 \si{Myr} (\textit{right}). \textit{Right panel:} distribution of kinetic energy density ($\rho {\rm v}^2$) in the $x$-$y$ plane at time 2.23 Myr showing a prominent S-shaped feature near centre. The values of unit pressure and unit kinetic energy density are $4.57 \times 10^{-12}$ $\rm dyn/cm^2$ and $4.57 \times 10^{-12}$ $\rm erg/cc$ respectively, and the lengths are defined with respect to the unit length 200 pc. The density distribution of the galaxy follows this pressure distribution. All shown distributions here are for \textit{CaseB}.} 
    \label{fig:jet_distribution}
\end{figure*}

\begin{figure}
    \centering
    \includegraphics[width=\columnwidth]{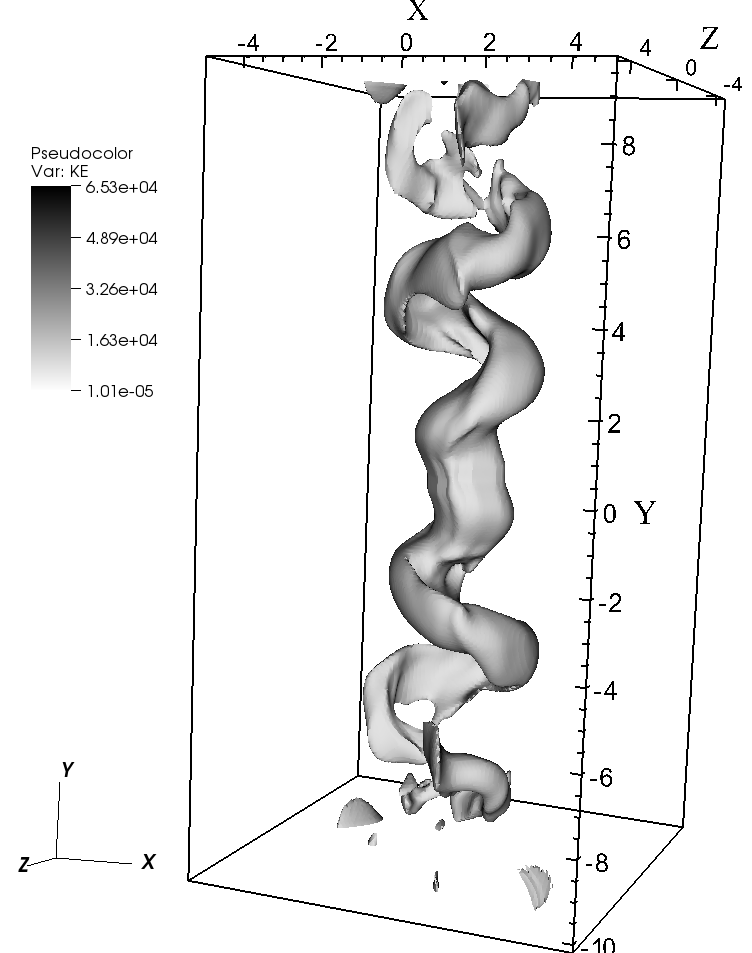}
    \caption{3D isosurface plot of the kinetic energy density of the synthetic jet structure, representing its helical motion at 2.23 Myr. The jet is losing its collimation as it propagates to longer distances, highlighting the transition zone, where it converts its helical motion into rather a continuous hollow non-precessing propagation. Here lengths and kinetic energy density distribution are defined with respect to 200 pc and $4.57 \times 10^{-12}$ $\rm erg/cc$ respectively.} 
    \label{fig:isosurface}
\end{figure}

The dynamical evolution of the precessing jet at simulation time 0.75 Myr, 1.5 Myr and 2.23 Myr is represented in Fig.~\ref{fig:jet_distribution} (\textit{left panel}) through the pressure distribution of the galaxy for \textit{CaseB}. In the initial phase, due to the precession, a helical structure of the jet starts to appear along with the presence of a prominent bow shock formed due to the jet-ambient medium interaction. As the jet evolves, the forward shock weakens significantly and the helical structure of the jet forms a distinct S-shape at the centre which is prominent in the kinetic energy plot shown in Fig.~\ref{fig:jet_distribution} (\textit{right panel}). This is what has actually been observed for the source 2MASX J1203 where no emission is found from the bow shock regions within the observed field of view \citepalias{rubinur2017}. From Fig.~\ref{fig:jet_distribution}, it is evident that the strength of the bow shock decreases with time and eventually the galaxy as a whole becomes a weak shock system. This fact is also verified from the compression ratio study of the injected Lagrangian particles, and at time 2.23 Myr, a majority of these particles were found to have undergone shocks having compression ratio $\sim 2$. For \textit{CaseA}, the jet is incapable of penetrating the circumjacent area as like in \textit{CaseB} due to a lower propagation speed of the jet. As a result, the formation of a similar structure (as Fig.~\ref{fig:jet_distribution}) occurs at a later time than \textit{CaseB}. Considering the distorted morphology of the jet, we further expanded the kinetic energy plot shown in Fig. \ref{fig:jet_distribution} in 3D to show an extended view of the helical motion of the jet. In Fig. \ref{fig:isosurface}, we illustrate it via the isosurface plot of kinetic energy density of our simulated structure drawn at time 2.23 Myr. We find that the jet is losing its collimation as it propagates to greater distances, forming a transition zone (after travelling $\sim$ 1.6 kpc) where the helical jet motion transitions into a continuous hollow non-precessing jet flow \citep[see][]{Nawaz2016}. This has also been reported in the simulation of \citet{Monceau2015} while modelling the precessional motion of SS~433, a famous X-ray binary in our galaxy. In the figure, the cocoon structure is not captured (as also in Fig. \ref{fig:jet_distribution}) as the kinetic energy of it is significantly lower than that of the jet. 

\begin{figure}
    \centering
    \includegraphics[width=\columnwidth]{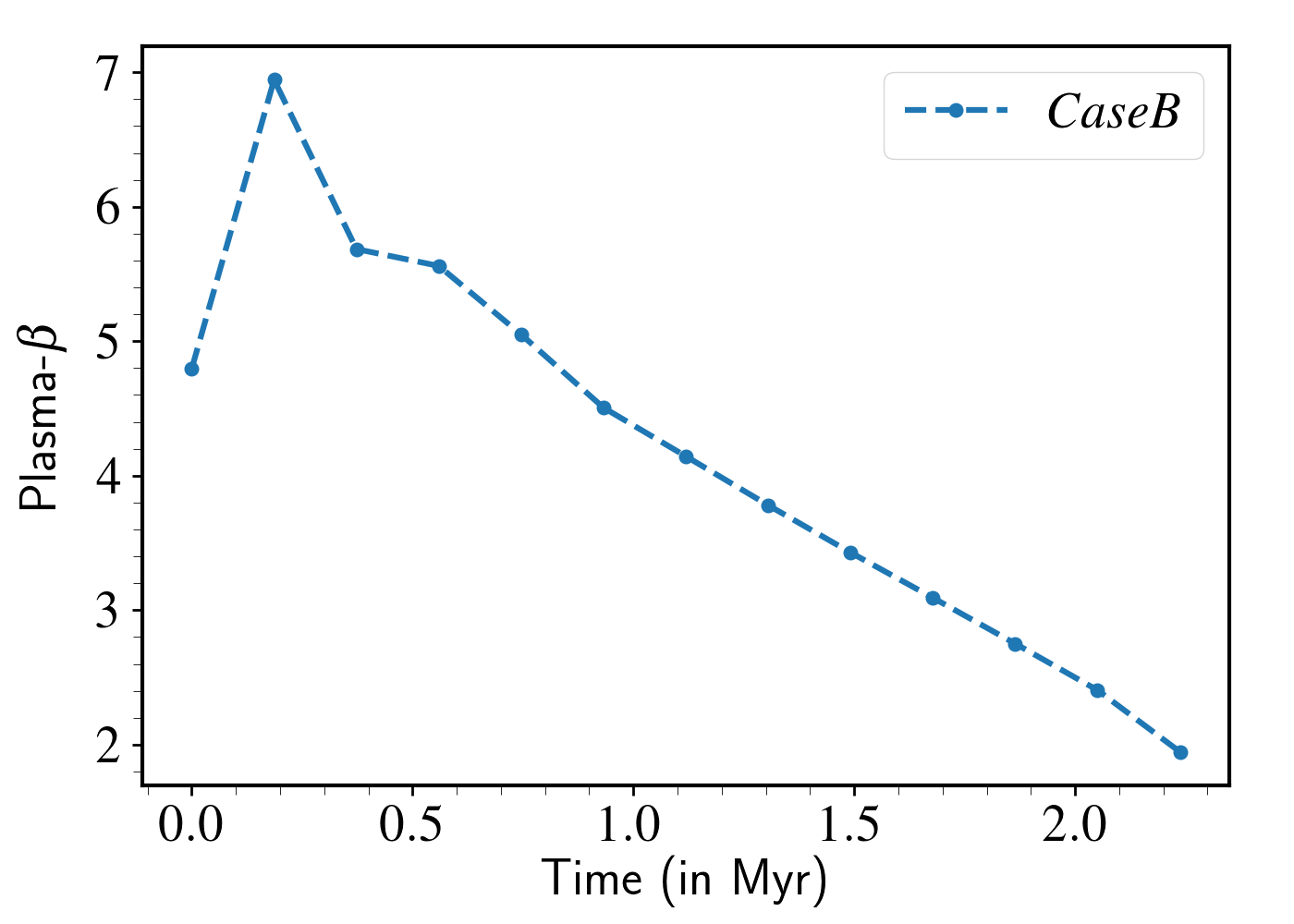}
    \caption{Time variation of plasma beta ($\beta$) showing its evolution as the structure grows for \textit{CaseB}. The $\beta$ values suggest that with time, magnetic and thermal energy of the cocoon reaches equilibrium.} 
    \label{fig:Beta}
\end{figure}

Using the ratio of thermal to magnetic pressure as $\frac{8\pi P}{B^2}$, we have evaluated the averaged value of plasma-$\beta$ of our simulated structure, and hence its evolution with time (Fig. \ref{fig:Beta}).
The plasma-$\beta$ values do not show substantial variation with time. Starting from a value of 4.8 at the beginning, it rises to 7 at time 0.37 Myr, before gradually decreasing to a value close to unity at time 2.23 Myr. This variation shows that the cocoon's magnetic and thermal energy approach equilibrium over time.
The density-weighted average magnetic field we obtained at time 2.23 Myr of our simulated structure is $\sim$ 60 $\mu$G, which is lower than the value suggested for 2MASX J1203 (i.e. 105 $\mu$G). However, one should note that the value of the magnetic field estimated in \citetalias{rubinur2017} is the equipartition field strength ($B_{\rm eq}$) which cannot be regarded as the dynamical magnetic field ($B_{\rm dyn}$) in case a significant deviation from the equipartition approximation (Section \ref{Emission model}) exists for the observed source. The discrepancy found in $B_{\rm eq}$ (observed) and $B_{\rm dyn}$ (estimated here) further showcases the fact that the galaxy 2MASX J1203 does not satisfy the equipartition condition (discussed further in Section \ref{Equipartition approximation and Age estimation}). For such a case, the value of $B_{\rm dyn}$ is typically found to be a few times lesser than $B_{\rm eq}$ as shown by \citet{Croston2005, mahatma2020}.

The effect of the initial magnetic field strength in the simulated structure is crucial to consider, as a higher value of toroidal field can suppress the lateral expansion of the precessing jet and hence the S-morphology. For a lower strength of the magnetic field, the field lines become incapable of constraining the plasma blob formation due to instabilities in the jet resulting in many knot-like features \citep{Bodo2013,Mukherjee2020,Borse2021}. Furthermore, the synchrotron cooling time increases as the magnetic field strength decreases \citep{Fan2008}, resulting in an increased emission from these generated blobs, reducing the prominence of the formed S-shape. Further discussion on this follows in Section \ref{sec: total intensity continuum} in terms of the synthetic emission maps. 

\begin{figure}
    \centering
    \includegraphics[width=\columnwidth]{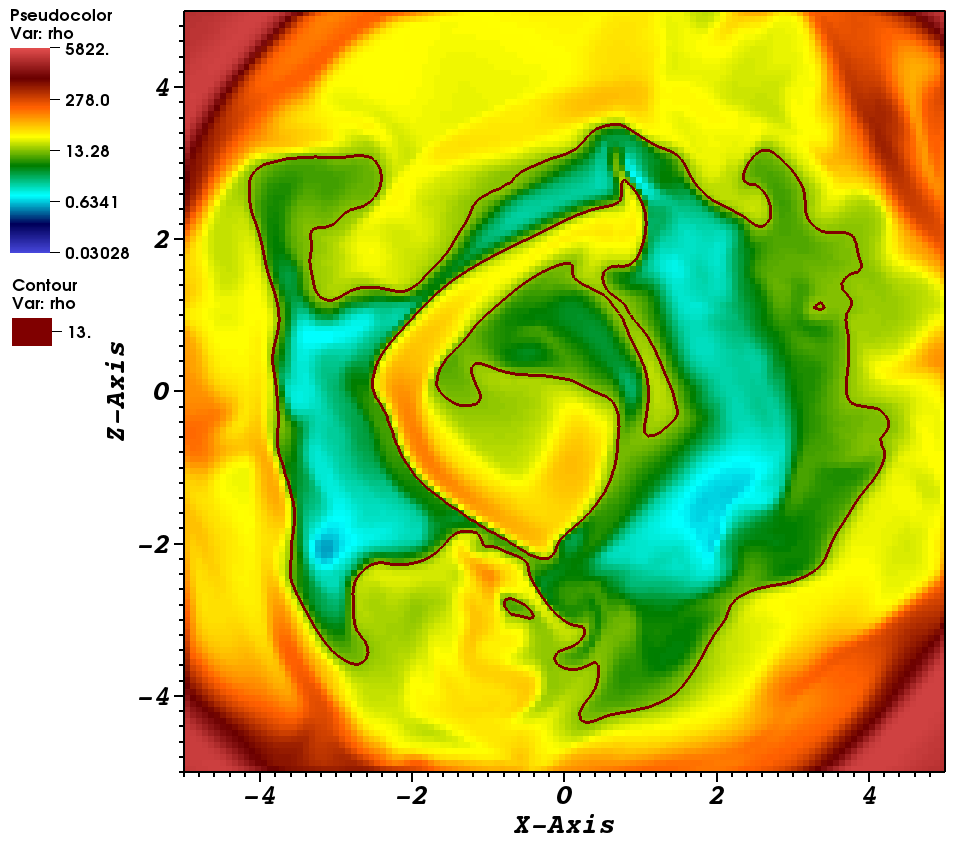}
    \caption{An $x$-$z$ slice at $y = 4$ (computational unit) of our simulated 3D run (\textit{CaseB}) showing the formation of inhomogeneous structures (at time 2.23 Myr) indicating the presence of KH instabilities. The inner contour prominently highlights the helix morphology of the jet, followed by the extended turbulent structure generated primarily due to KH instability. The lengths and density defined here are with respect to 200 pc and $10^{-3}$amu/cc, respectively.} 
    \label{fig:instabilities}
\end{figure}

An analysis of the Alfv\'enic Mach number ($\mathcal{M}_A = {\rm v}_{\rm jet}/{\rm v}_A$; ${\rm v}_A = B/\sqrt{4\pi\rho}$) of the jet shows that it has been injected into the domain initially with a super  Alfv\'enic speed ($\mathcal{M}_A \sim 6.5$). This is because our injected jet is kinetically dominated (Section \ref{sec:dynsetup}), which is also the regime for Kelvin-Helmholtz (KH) instabilities to grow.  As a result of which, the jet  experiences an additional deceleration, and we see a reduction in values of $\mathcal{M}_A$ reaching to 1.14 at 2.23 Myr. Such trans Alfv\'enic systems are prone to have instabilities of both the Kink and KH modes \citep[see][]{Acharya2021} which we have also observed in our case (i.e. for \textit{CaseB}). In Fig. \ref{fig:instabilities}, we show a 2D cut of density distribution at 2.23 Myr in the $x$-$z$ plane (perpendicular to the precession axis) at $y = 4$ (computational unit) showing the evidence of decollimated turbulent flow originated due to KH instabilities. The contours clearly highlight the helix of the jet followed by the inhomogeneous extended structures generated as the result of KH instabilities \citep[similar to][]{Baty2002,Monceau2015,Borse2021}. On the other hand, twisting and bending of jet or inhomogeneous expansion of cocoon are thought to be the signatures of kink instabilities as found in low powered jet \citep{Mukherjee2020}. In our case, we observed that the overall cocoon geometry gets curved as time progresses, a hint of which can be seen in Fig. \ref{fig:jet_distribution} (middle panel of pressure distribution plot). The asymmetric expansion of the cocoon is visible in the figure in regions $|y| \gtrsim 5$ indicating a possible presence of kink instabilities in these low powered jetted systems. 
The point we want to highlight here is that these MHD instabilities play their part in slowing down the jet, contributing to the reduction of average jet kinetic energy values with time.

\section{Synthetic Spectral Characteristics} \label{Spectral signatures}
Here, we discuss the emission characteristics of the synthetic structures obtained from our simulations.
\subsection{Total Intensity Continuum} \label{sec: total intensity continuum}
\begin{figure*}
    \centering
    \includegraphics[width=2.05\columnwidth]{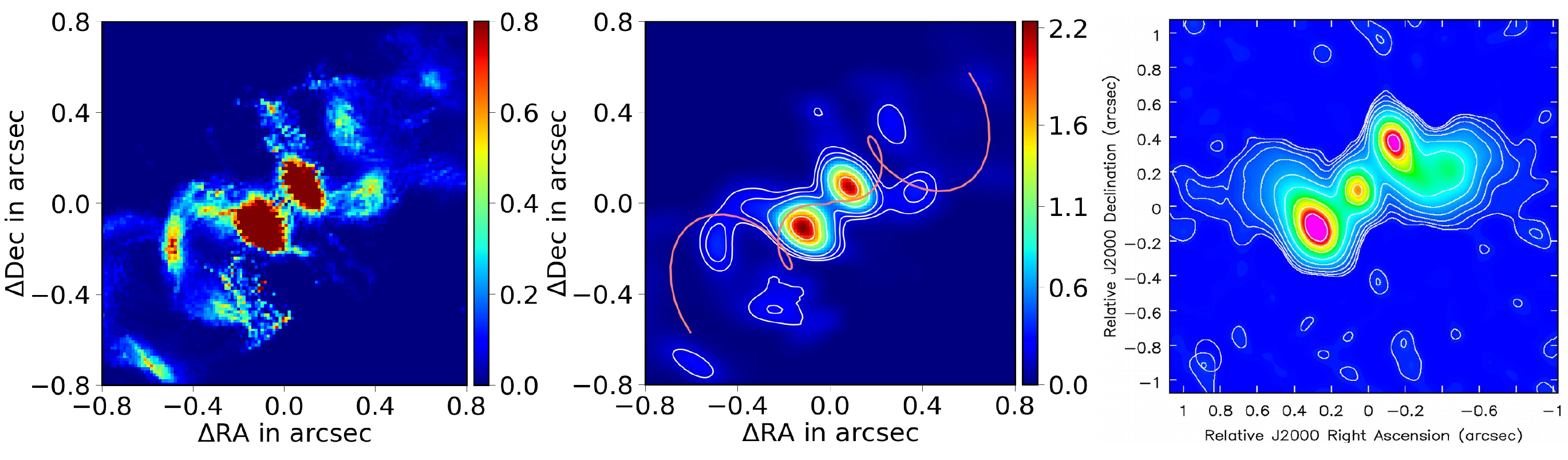}
    \caption{\textit{Left:} synthetic high resolution intensity map obtained for \textit{CaseA} at 11.5 GHz at time 2.98 Myr. \textit{Centre:} the same intensity map as the \textit{left} one, but convolved to the resolution as mentioned in \citetalias{rubinur2017} with contour levels at 8, 12, 20, 30, 40, 60, 80 percent of peak flux value. We have additionally over-plotted here a modelled helical jet trajectory (the magenta curve) obtained using the \citetalias{hjellming1981} model with parameters from \citetalias{rubinur2017}, implying that our simulated structure appears compressed in size when compared to the astronomical source (see Section \ref{Spectral signatures} for details). \textit{Right:} the observed emission map of 2MASX J1203 at the same frequency, plotted in a relative scale for comparison. In the synthetic maps, the emission from the core is missing as we have not explicitly modelled the core region.}
    \label{fig:11.5GHz}
\end{figure*}

For the spectral morphology of the galaxy, let us first consider \textit{CaseA} where the parameters used for the simulation are obtained from \citetalias{rubinur2017}. We have presented in Fig. \ref{fig:11.5GHz}, the high resolution synthetic image of the galaxy at 11.5 GHz at time 2.98 Myr (\textit{left}), its Gaussian convoluted version with beam size $0.19\arcsec \times 0.13\arcsec$ as mentioned in \citetalias{rubinur2017} (\textit{middle}) and the observed map of 2MASX J1203 at the same frequency (\textit{right}).
Our synthetic map shows resemblances with the observed radio image of 2MASX J1203 in terms of appearance, leading to the fact that a precessing jet can produce the observed morphology of S-shaped winged sources. 
The helical jet features emerging from the distinct hotspots are prominent in the simulated map. Since we do not model or introduce an ad hoc core region explicitly, the emission from the central core is not captured in our maps. However in terms of the physical size, our obtained structure decelerates significantly. The galaxy as a whole appears a bit compressed in size compared to the observed image shown in Fig. \ref{fig:11.5GHz}. This is further verified from the obtained hotspot distance of $0.28 \arcsec$ for our case in comparison with $(0.6\pm 0.1)\arcsec$ for the observed source.
To better understand the discrepancy between the observed source size and simulated size, we have over-plotted a model of a helical jet trajectory (in magenta color) obtained using \citetalias{hjellming1981} model with parameters provided by \citetalias{rubinur2017} for the source 2MASX J1203. Despite the use of the same parameters in our simulation and in the modelled jet trajectory, the sign of deceleration is obvious in Fig. \ref{fig:11.5GHz}. 
This is expected as the \citetalias{hjellming1981} model does not consider the effect of interaction between the jet and the ambient medium that plays a crucial role in constraining the free expansion of the jet \citep{Panferov2014}. Furthermore, the deceleration rises owing to the jet's helical motion \citep{Monceau2014}, and additional deceleration comes from MHD jet instabilities as shown in Section \ref{sec:dynamics}. These effects we have obtained self-consistently in our simulations. So, we conclude here that the parameters estimated for a precessing source by tracking the jet locus using \citetalias{hjellming1981} model are not sufficient and are underestimated.

\begin{figure}
    \centering
    \includegraphics[width=\columnwidth]{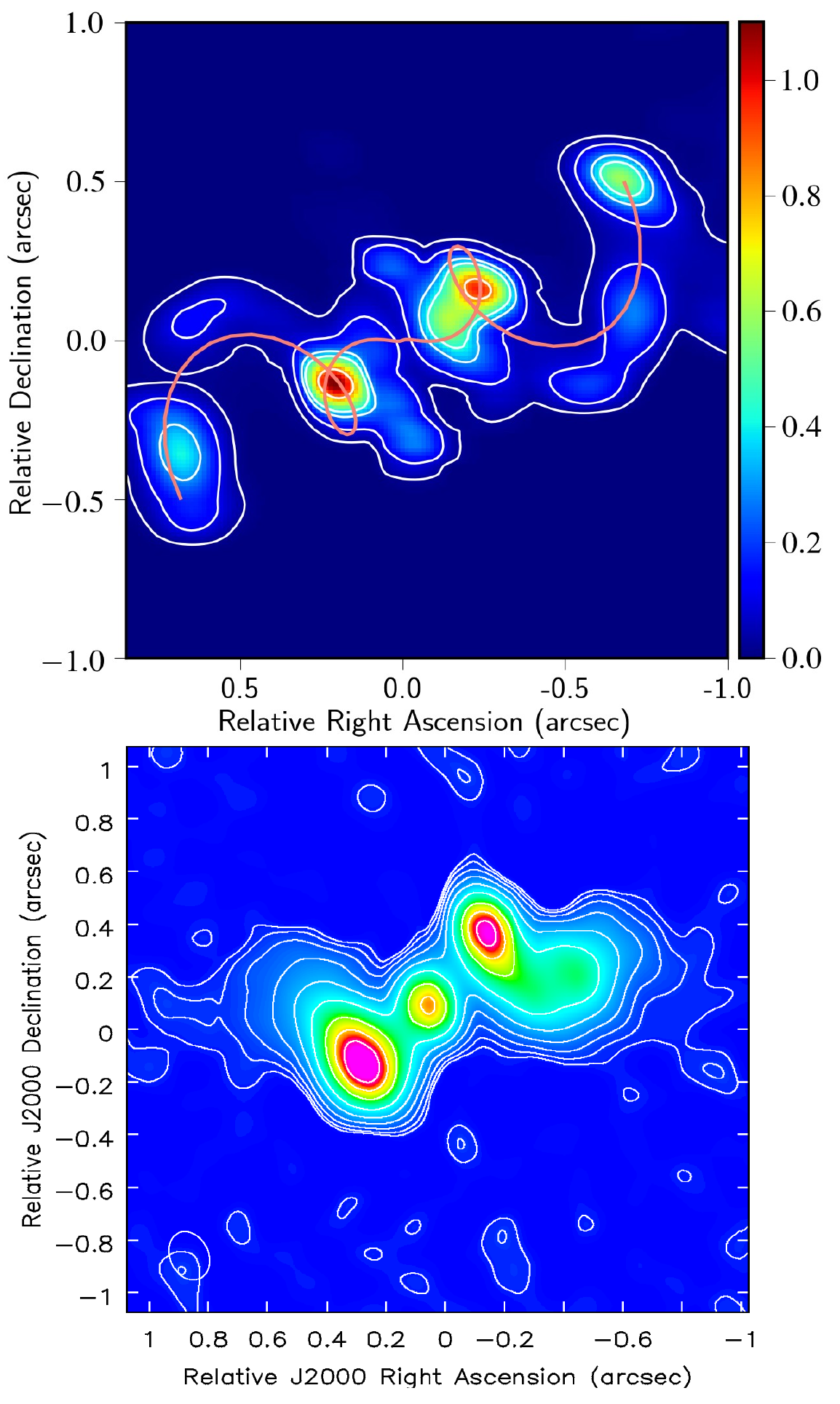}
    \caption{\textit{Top:} synthetic intensity map obtained for \textit{CaseB} at 11.5 GHz at time 2.23 Myr, convolved to the resolution mentioned in \citetalias{rubinur2017}. The contour levels are drawn at 0.6, 2.5, 7, 10, 20, 40, 60, 80 percent of peak flux value. The formation of two distinct hotspots near centre with an S-shape morphology are also prominent in this case. The over-plotted magenta curve represents the modelled helical jet arm obtained using \citetalias{hjellming1981} model that typically represents the observed source. \textit{Bottom:} the observed emission map of 2MASX J1203 (at 11.5 GHz), shown in the same relative scale for comparison. Here, the contours are put at similar locations as mentioned in \citetalias{rubinur2017}. The synthetic structure resembles well with the observed source both in terms of their size and shape (see section \ref{sec: total intensity continuum} for details).}
    \label{fig:11p5GHz_CaseB}
\end{figure}

To counter the effect of deceleration discussed above, we have considered \textit{CaseB} where the jet velocity is increased by 1.25 times with a slight adjustment in other parameters as shown in Table \ref{tab:simsetups}. The result obtained from this run is presented in Fig. \ref{fig:11p5GHz_CaseB}. The convoluted intensity map at 11.5 GHz (at time 2.23 Myr) is shown in the \textit{top} where the observed image of 2MASX J1203 at the same frequency is shown in the \textit{bottom} in the same relative scale for comparison. Fig. \ref{fig:11p5GHz_CaseB} evidently shows that the size of our simulated structure has increased significantly in comparison to \textit{CaseA} (at a lower time as well) due to the adjustment in jet parameters. This implies that the resulting structure is sensitive to parameter choices, which we have also verified and highlighted further in this section. The hotspot distance obtained for this case is $0.543\arcsec$ which is comparable to the distance of ($0.6\pm0.1$)$\arcsec$ for the observed source. In terms of morphology, the structures are also quite alike. The presence of two distinct hotspots near the centre (north-west (NW) and south-east (SE) as per \citetalias{rubinur2017}) with helical arms emerging from it are prominent in the synthetic image as also has been observed for 2MASX J1203. Especially, the diverted flow associated with the NW hotspot (near $\Delta$RA, $\Delta$Dec: -0.3, 0.1) is identical to the bent jet feature seen in the source 2MASX J1203 at a similar location. The formation of two secondary knots is observed in our synthetic map at locations ($\Delta$RA, $\Delta$Dec) as (-0.7, 0.5) and (0.7,-0.4) whose relevance in the observed map is not very obvious. Although the extended structures emerging from these secondary hotspots (especially at the extreme right) resemble the structures captured by the observed source's outermost contour, we shouldn't attach too much emphasis on them since they are low surface brightness extensions of the dominant S-shape. The cavity observed towards the south of the main structure in the synthetic image is a result of the missing core that we did not explicitly include in our simulations. In the presence of a compact core, the back-flowing plasma formed in the jet will be deflected strongly in the lateral direction ensuring that the cavity region is steadily filled. Further, to understand the resemblance between the observed and simulated structure better, we have over-plotted a modelled helical jet trajectory (in magenta color) obtained using \citetalias{hjellming1981} model with parameters provided by \citetalias{rubinur2017} over our synthetic intensity image (see Fig. \ref{fig:11p5GHz_CaseB}). To be specific, the parameters used for the \citetalias{hjellming1981} model are ${\rm v}_j = 0.023c$, $\psi = 23^{\circ}$, $\tau = 10^5$ yrs and ($\theta,\ \upvarphi,\ \chi$) as ($90^{\circ},\ 42^{\circ},\ 30^{\circ}$). These values are within the error limit of parameters for the astronomical source 2MASX J1203 as suggested by \citetalias{rubinur2017}. We see that the magenta curve tracks well the structure that are obtained from our simulation (for \textit{CaseB}) with parameters ${\rm v}_j = 1.25 \times 0.023c$, $\psi = 25^{\circ}$, $\tau = 0.95 \times 10^5$ yrs and ($\theta,\ \upvarphi,\ \chi$) as ($90^{\circ},\ 20^{\circ},\ 19^{\circ}$). From this discussion, we can see that the velocity used in the modelled curve and in the simulation is not same, and the magenta curve best fits a simulated structure that has been produced with a higher jet velocity. The additional changes observed in other parameters can also be attributed to the effect of decelerations that the jets have encountered self consistently in our simulations. This further supports the result reached by \citet{Monceau2014} for the micro-quasar SS 433, namely that the deceleration effects encountered by the jet must be accounted for in the kinematic models to predict the precession parameters correctly.

\begin{figure*}
    \centering
    \includegraphics[width=2.05\columnwidth]{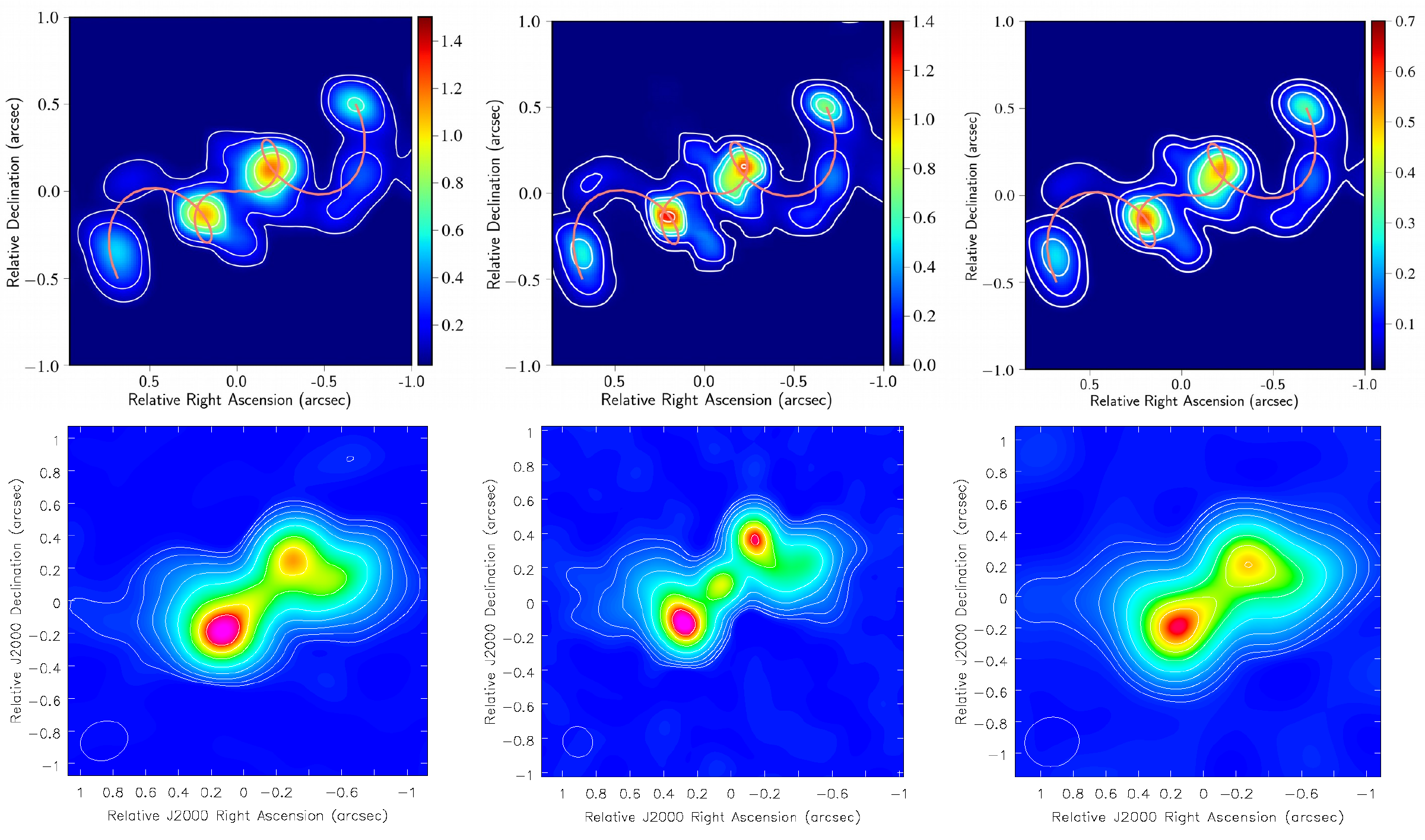}
    \caption{\textit{Top panel:} synthetic emission maps obtained for \textit{CaseB} at frequencies 6 GHz ( \textit{top left}), 8.5 GHz ( \textit{top centre}) and 15 GHz ( \textit{top right}) at time 2.23 \si{Myr}. These maps are convolved to the resolutions as mentioned in \citetalias{rubinur2017}. The contours in the 6 \si{GHz} map correspond to 1.1, 2.5, 7, 10, 20, 40, 60 and 80 percent of the maximum flux value obtained at that frequency. Similarly, for 8.5 \si{GHz} map, the contours are drawn at 1.0, 2.5, 5.5, 10, 20, 40, 60 and 80 percent and for the 15 GHz map, it is drawn at 1, 2, 5.5, 8, 12, 20, 40, 60 and 80 percent of the maximum flux values respectively. \textit{Bottom panel:} observed emission maps of 2MASX J1203 for the same frequencies (i.e. 6 GHz (\textit{left}), 8.5 GHz (\textit{centre}) and 15 GHz (\textit{right})). Despite the formation of a few secondary features, the central morphology appears to be similar for both of our synthetic and observed maps, as has been demonstrated by the magenta lines drawn using \citetalias{hjellming1981} model.}
    \label{fig:rest_emission}
\end{figure*}

The synthetic emission maps for other observing frequencies i.e. at 6 GHz, 8.5 GHz and 15 GHz, generated from \textit{CaseB} at time 2.23 Myr, are shown in Fig. \ref{fig:rest_emission} (\textit{top row}). They are convolved to the resolutions mentioned in \citetalias{rubinur2017}. In the \textit{bottom panel} of Fig. \ref{fig:rest_emission}, we present the observed maps of 2MASX J1203 at the same frequency bands for comparison. Despite the appearance of a few additional sub-structures (secondary blobs) in our emission maps, the central morphology (i.e. dominant hotspots with emerging arms) is found to be similar for both the cases. This can also be inferred from the over-plotted magenta colored lines drawn using \citetalias{hjellming1981} model (Fig. \ref{fig:rest_emission}). The parameters used for the lines are same as used in Fig. \ref{fig:11p5GHz_CaseB}.

We have calculated the flux value of our synthetic radio structure at 11.5 GHz as 0.21 Jy (at time 2.23 Myr for \textit{CaseB}) which is similar to the measured value of 0.29 Jy of 2MASX J1203. Additionally, we have estimated the peak flux values of SE and NW hotspots at the same observing frequency as 3.44 mJy and 3.02 mJy respectively which is similar to the observed value of 4.90 mJy and 4.44 mJy respectively. So, the ratio of peak SE hotspot flux ($f_{\rm{SE}}$) to the NW hotspot flux ($f_{\rm{NW}}$) becomes 1.14 at 11.5 GHz for our case. Accordingly, the values of $\frac{f_{\rm{SE}}}{f_{\rm{NW}}}$ for 6 GHz, 8.5 GHz and 15 GHz are obtained as 0.96, 1.14 and 1.13 respectively which are in line with the obtained values for 2MASX J1203. We note here that this formed radio structure has started to evolve from an initial under-equipartition condition where radiating electrons energy was a fraction of the magnetic energy of the jet (see Section \ref{Emission model}). In this context for the run \textit{CaseB\_eqp}, the flux values computed at the hotspots are $\sim 15$ times higher and of the whole galaxy, $\sim 25$ times higher than the value specified in \citetalias{rubinur2017}. This is due to the fact that we have started with a greater non-thermal electron density in the jet (Table \ref{tab:simsetups}) which was computed assuming equipartition between the radiating electrons energy and the jet's magnetic energy.
This further hints at the existence of an initial under-equipartition state of these galaxies, which is used in \textit{CaseB} yielding the best match between the observed and simulated structure. The time evolution of this initial condition is further highlighted in Section \ref{Equipartition approximation and Age estimation}. 

\begin{figure*}
    \centering
    \includegraphics[width=2\columnwidth]{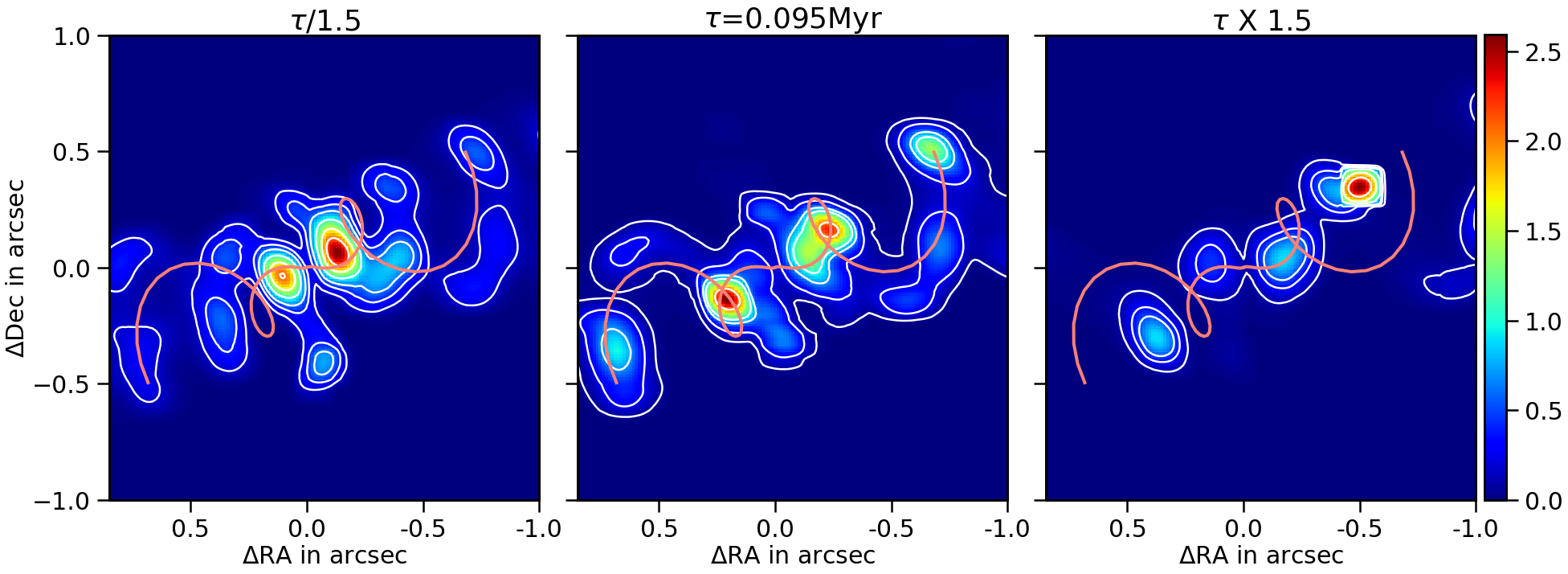}
    \caption{Synthetic 11.5 GHz intensity maps showing resultant morphologies, obtained from our simulations with different jet precession period ($\tau$) choices (at 2.23 Myr). A precession period of 0.095 Myr as suggested for the astronomical source 2MASX J1203, has been decreased by 1.5 times in the \textit{left} image (\textit{CaseB\_lPP}), kept the same in the \textit{middle} image (\textit{CaseB}) and has been increased by 1.5 times in the \textit{right} image (\textit{CaseB\_hPP}). The images further suggest that the formation of S-morphology is sensitive on the parameter selections. The modelled jet arms in magenta color are also over-plotted here to demonstrate how much the formed structures deviate or resemble in morphology compared to the observed source.}
    \label{fig:pp}
\end{figure*}

As previously indicated, the parameter selections affect the resulting structure. Here, we demonstrate how morphological changes appear by our modest change in the precession period ($\tau$). In Fig. \ref{fig:pp}, we show synthetic intensity images of three structures that are obtained at time 2.23 Myr for \textit{CaseB\_lPP} (\textit{left}), \textit{CaseB} (\textit{middle}) and \textit{CaseB\_hPP} (\textit{right}). For these cases, we have only changed the jet precession period by 1.5 times lower, kept the same and 1.5 times higher than the period estimated for the source 2MASX J1203 respectively (see Table \ref{tab:simsetups}). We can see from Fig. \ref{fig:pp} that for the case with a smaller precession period $\tau$, the jet rotates faster resulting in a compact S-structure. Whereas, with increasing $\tau$ the structure loosens up \citep[similar to][]{Hardee2001}. For \textit{CaseB\_hPP}, $\tau$ is so large that the resulting morphology loses its S-shape completely. Despite the fact that we only changed $\tau$ by a factor of 1.5, it is still capable of completely changing the size and geometry of the produced structure. This suggests that parameter selection has an impact on S-morphology formation and in this regard, numerical simulation becomes an efficient way of constraining the parameter values.

\begin{figure}
    \centering
    \includegraphics[width=\columnwidth]{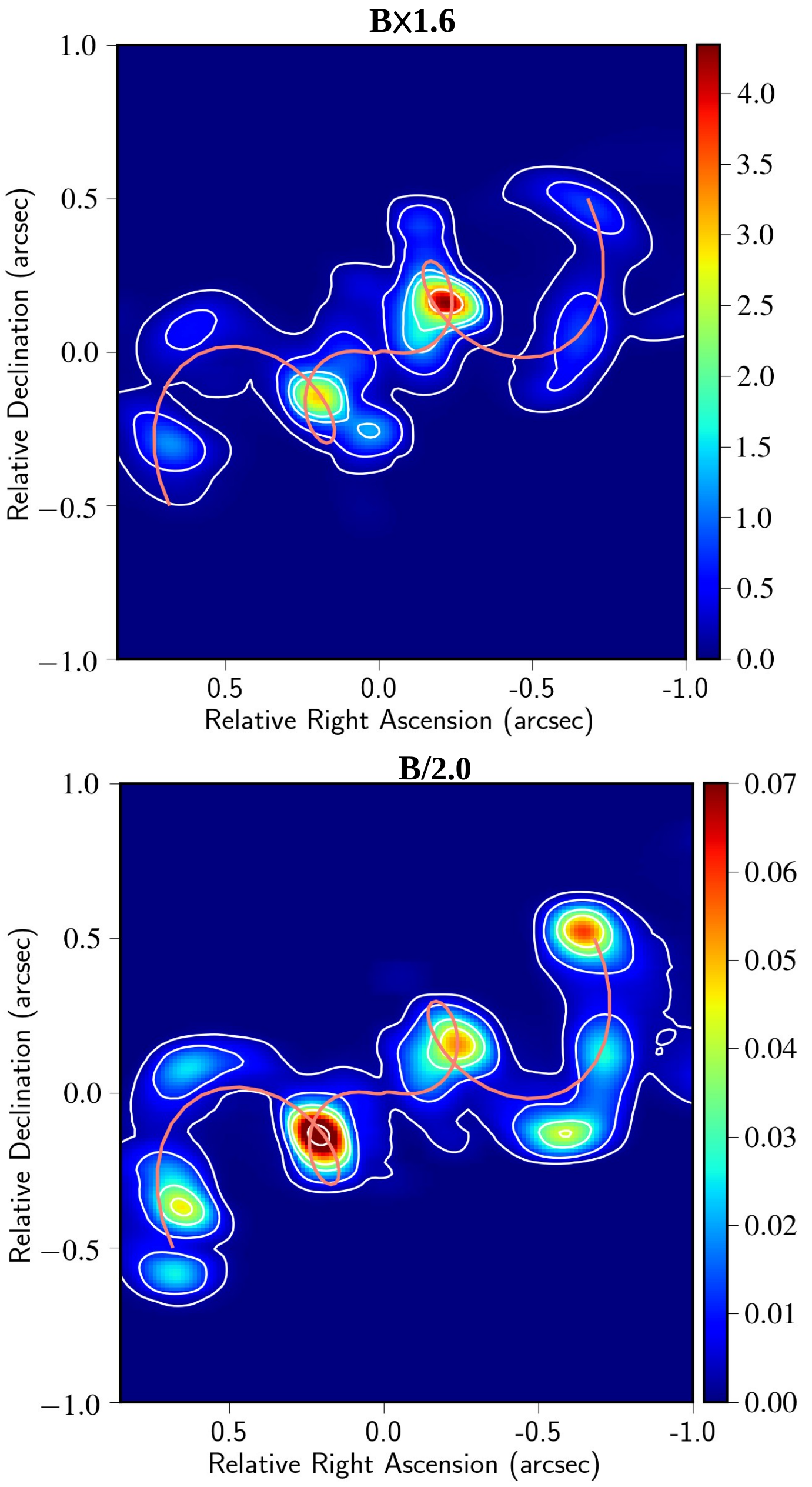}
    \caption{Here, we show the changes that appear in the jet morphology at a time of 2.23 Myr as a result of changing the injected jet magnetic field strength. The intensity maps at the top and the bottom are for \textit{CaseB\_hB} and \textit{CaseB\_lB} respectively. The contours in both the maps are put at same levels as in Fig \ref{fig:11p5GHz_CaseB} for comparing the morphologies with \textit{CaseB}, which is an intermediate B-field strength run (i.e., B = 0.19 mG). See Section \ref{sec: total intensity continuum} for more details.}
    \label{fig:hlB}
\end{figure}

To strengthen the above mentioned fact, we also showcase here the results obtained from our modified magnetic field runs, i.e., from \textit{CaseB\_hB} with higher magnetic field and from \textit{CaseB\_lB} with lower magnetic field (compared to \textit{CaseB}; Table \ref{tab:simsetups}). The results are shown in Fig. \ref{fig:hlB}, highlighting distorted morphologies. We infer from Fig. \ref{fig:hlB} that with increasing magnetic field strength, the helical arms of the structure tighten up, creating a tightly wrapped structure. As also discussed in Section \ref{sec:dynamics}, for the lower B-field case, we see the presence of multiple bright knots in the jet arms (even intensity of some secondary hotspots is higher than the primary ones), contrary to what has been observed for 2MASX J1203 owing to the effects of instabilities \citep[primarily KH,][]{Bodo2013} and higher synchrotron cooling time for this case. We thus note that increasing or decreasing the magnetic field strength also has a major impact on the structure, modifying its morphology considerably than the one obtained in \textit{CaseB}. 

\subsection{Spectral index distribution} \label{Spectral index distribution}
\begin{figure}
    \centering
    \includegraphics[width=1\columnwidth]{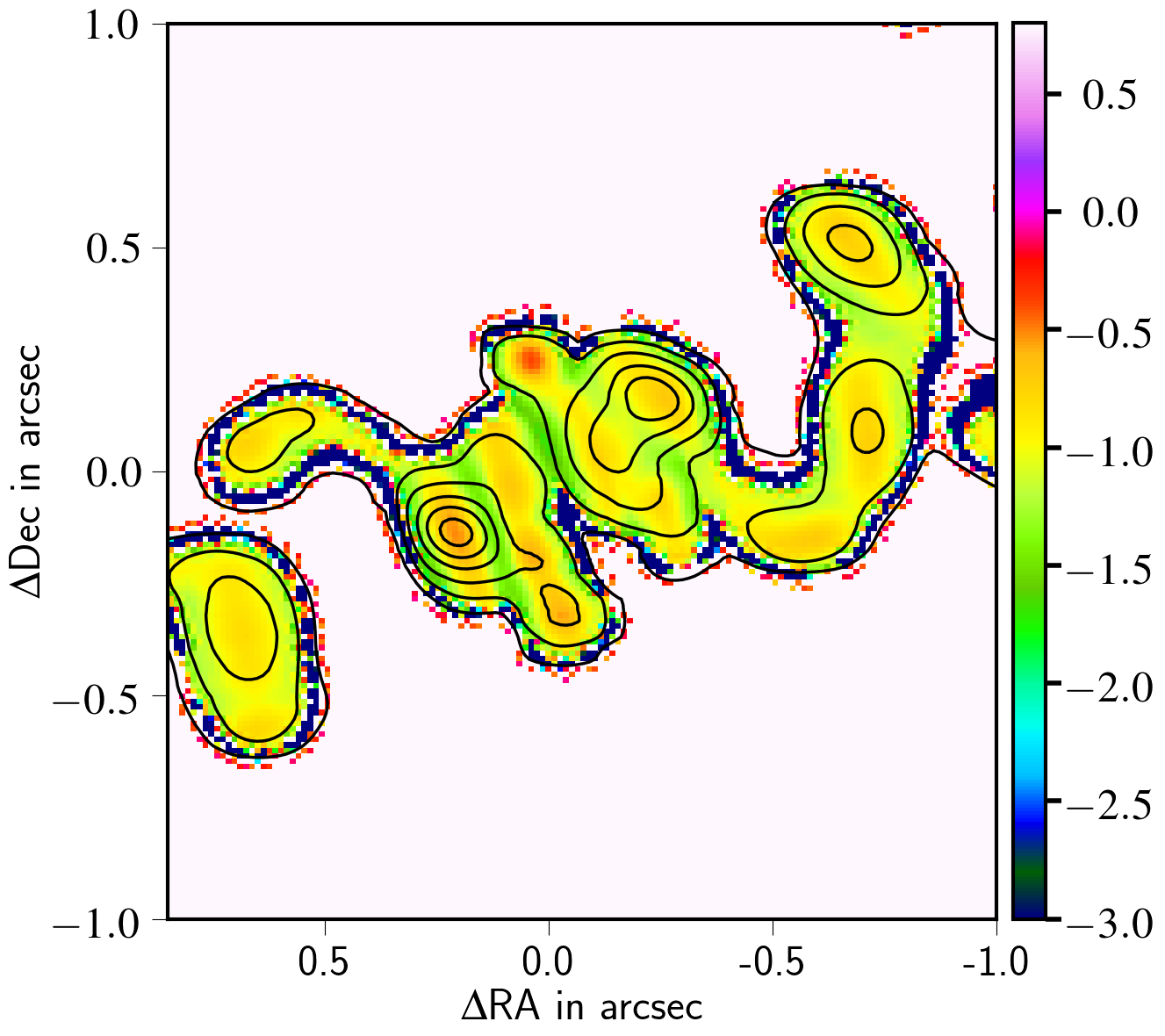}
    \caption{Spectral index ($\alpha$) map of our obtained structure (from \textit{CaseB} at time 2.23 Myr) for frequencies 8.5-11.5 GHz. We see the steepening of $\alpha$ values in the side and leading edges of the jet arm. Whereas, the $\alpha$ values are relatively flatter along the jet ridge lines. The core region is showing steep $\alpha$ here, due to the lack of an ad-hoc core that we have not explicitly considered. The hotspots show flat $\alpha$ values with $\alpha \sim - 0.75$ (see Section \ref{Spectral index distribution}).}
    \label{fig:alpha}
\end{figure}

As per the synchrotron radiation, the flux density $S_{\nu} \propto \nu^{\alpha}$, where $\nu$ is the frequency of observation and $\alpha$ is the spectral index. The value of $\alpha$ can be obtained as 
\begin{equation}
    \alpha = \rm{log}_{10}(S_{\nu1}/S_{\nu2})/\rm{log}_{10}(\nu1/\nu2)
\end{equation}
Its distribution for \textit{CaseB} for the observing frequencies of $\nu1 = 8.5$ GHz and $\nu2 = 11.5$ GHz is shown in Fig. \ref{fig:alpha}. The obtained spectral index map possess three distinct characteristics - a) spectral index values in the hotspots and compact knots (or humps) associated with the hotspots are flat ($\alpha \gtrsim -0.8$), b) spectral index values along the jet ridge lines are moderately steep ($-1 \lesssim \alpha \lesssim -0.8 $) and c) spectral index values in the jet side and leading edges are steep ($\alpha \lesssim -1$). These are also typically the characteristics observed in the $\alpha$-map of 2MASX J1203 obtained for the same observing frequencies \citepalias{rubinur2017}. One must note here that near the central region of our simulated structure, the $\alpha$-values are steep (contrary to the observed map) due to the lack of an ad-hoc core that we have not explicitly considered in our study. 

Further, we have evaluated intensity weighted average spectral index values as
\begin{equation}
    \alpha_{\rm av} = \frac{\int I_{\nu}(x, y)\alpha(x,y)dxdy}{\int I_{\nu}(x,y)dxdy}
\end{equation}
The values of $\alpha_{\rm av}$ (for $\nu1 = 8.5$ GHz, $\nu2 = 11.5$ GHz) obtained in the SE and NW hotspots are $-0.70$ and $-0.82$ respectively. This is relevant to the values highlighted by \citetalias{rubinur2017} for the same hotspots that corresponds to $-0.76$ and $-0.72$ respectively. In this regard, the run \textit{CaseB\_p4} is incapable of producing the desired $\alpha$-map which generates steeper $\alpha$-distribution throughout the galaxy. This further justifies our initial choice of a flatter power-law distribution of non-thermal electrons (i.e. $p = 3$) for these Seyfert galaxies.

\subsection{Polarization} \label{Polarization}
\begin{figure}
    \centering
    \includegraphics[width=1\columnwidth]{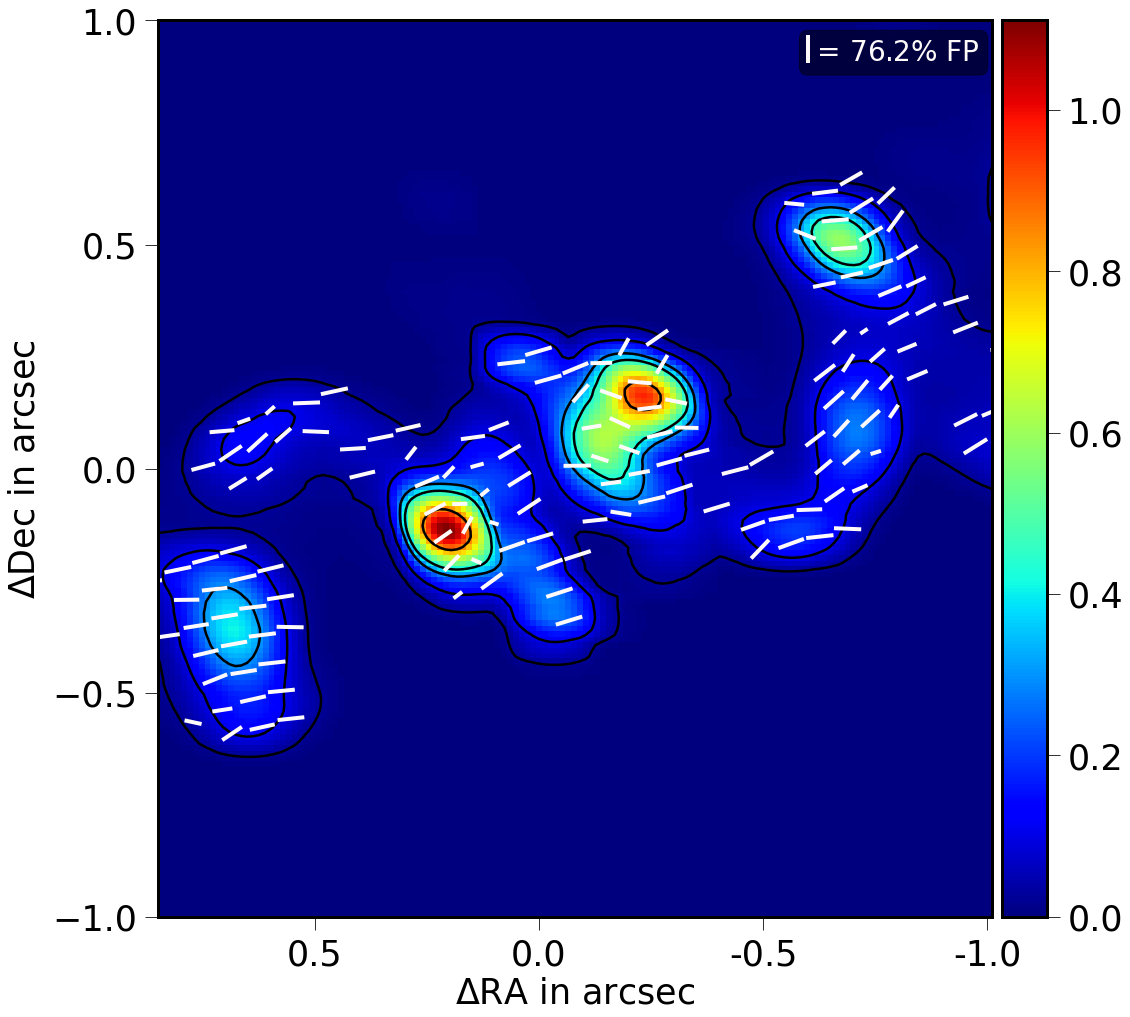}
    \caption{Polarization map (at 11.5 GHz) showing the magnetic field line distribution (white lines), plotted over the 11.5 GHz intensity structure. For better visualization and future comparison with observations, we have over-plotted the field lines over the rotated and convoluted intensity map. The size of the magnetic field lines are proportional to the fractional polarization (FP) at that point. In the top left corner, we show the scale length for maximum fractional polarization (76.2\%) obtained for the structure. The corresponding colorbar represents the intensity measure in Jy/arcsec$^2$.}
    \label{fig:polarization}
\end{figure}

We have also constructed the polarization map of our simulated galaxy at 11.5 GHz ($\nu$) from the linearly polarized synchrotron emission. The orientation of magnetic field lines is shown in Fig. \ref{fig:polarization} which is overlaid over the 11.5 GHz intensity map of the radio structure. 
For better visualization of the configuration and for future comparison with observations, we have over-plotted the field lines over the convolved and rotated intensity map of the galaxy.
The direction of magnetic field vectors are obtained by estimating the orientation of electric field vectors ($\Theta = 0.5 \tan^{-1} (U_{\nu}/Q_{\nu})$) and then rotating them by $90^{\circ}$. The length of the lines drawn in the image is proportional to the amount of fractional polarization at that point ($\propto \sqrt{U_{\nu}^2 + Q_{\nu}^2}/I_{\nu}$). Despite our incorporation of a toroidal magnetic field in the jet base, the observed B-field vectors along the chosen line of sight are parallel to the jet flow (Fig. \ref{fig:polarization}). Even at the hotspots where the jet is making a turn into another loop due to precession, the field lines are also observed to bend with the flow. The distribution shown in Fig. \ref{fig:polarization} is represented without accounting for the effect of Faraday Rotation.
We have found that the jet shows highly linearly polarized regions, mostly in the side and leading edges of the helical jet, ranging up to 76\%. This is consistent with the model where shock compression (formed due to the jet-ambient medium interaction) plays a crucial role in ordering the B-field and making them longitudinal \citep{Laing1981,Roberts2008}. These places are associated with moderately strong shocks having shock compression ratio $\sim 3$. Our result is also in line with the findings of \citet{Vaidya_2018} where they explicitly showed the presence of highly polarized regions in the shear interface between the jet and ambient medium (a location with strong oblique shocks). 

\begin{figure*}
    \centering
    \includegraphics[width=2\columnwidth]{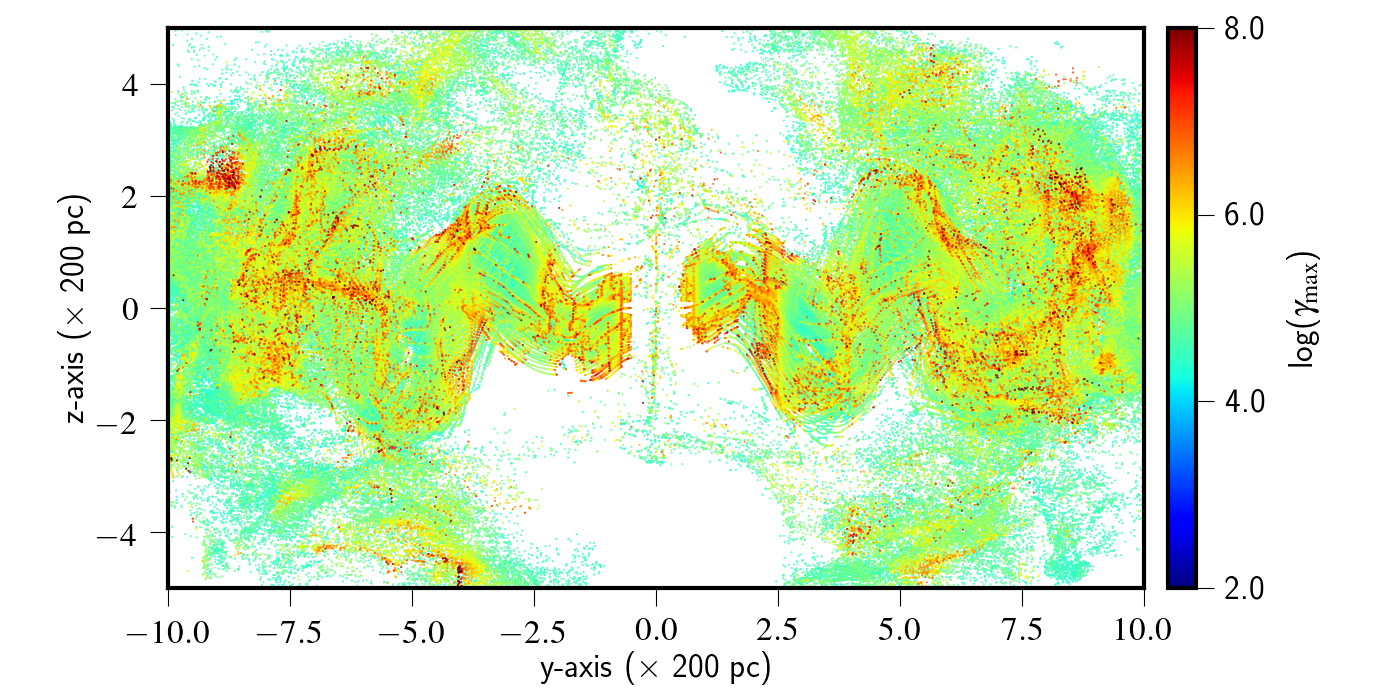}
    \caption{Particle distribution showing maximum Lorentz factor ($\gamma_{\rm max}$) of non-thermal electrons at a time when the structure has evolved for 2.23 Myr. This projected map (projected on the $y-z$ plane) shows evidence of re-energization (due to diffusive shocks) mostly on the side and leading edges of the jet forming high $\gamma_{\max}$ regions there. This further have implications on polarization map of the galaxy in ordering the B-field lines along the jet flow (discussed further in Section \ref{Polarization}).}
    \label{fig:gamma_max}
\end{figure*}

Further, the presence of these shocked regions can also be understood from Fig. \ref{fig:gamma_max} where we show the distribution of maximum Lorentz factor ($\gamma_{\rm max}$) values of the injected electrons. The distribution is projected on the $y-z$ plane for better visualization. From Section \ref{Emission model}, we know that we have started with an initial $\gamma_{\rm max}$ value of $10^{10}$ which will be reduced subsequently as it undergoes radiative and adiabatic cooling that we have considered in our simulations. However, these reduced values can increase significantly if they undergo strong shocks in their trajectory \citep{Borse2021,Mukherjee2021}. The distribution presented in Fig. \ref{fig:gamma_max} shows that despite the reduction of $\gamma_{\rm max}$ values across the domain, the side and leading edges of the jet show the presence of higher $\gamma_{\rm max}$ patches having values $\sim 10^{7}$ to $10^8$. These are mainly generated by moderately strong shocks having shock compression ratios $\sim 3$. The enhanced interaction between the helical jet and the ambient medium is primarily responsible for the shocks.

\subsection{Influence of diffusive shocks on particle spectra} \label{Influence of diffusive shocks on particle spectra}
\begin{figure*}
    \centering
    \includegraphics[width=2\columnwidth]{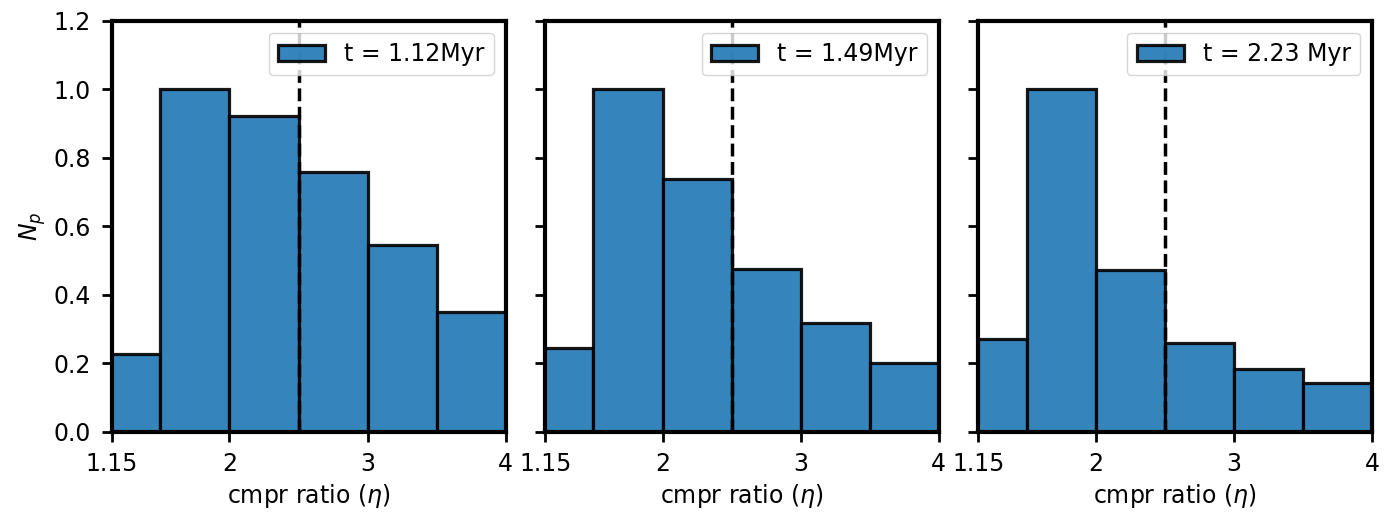}
    \caption{Histograms showing the distribution of compression ratio ($\eta$) of injected Lagrangian particles that are roaming in the computational domain at time 1.12 Myr (\textit{left}), 1.49 Myr (\textit{middle}) and 2.23 Myr (\textit{right}). The vertical axis represents the normalized number of particles ($N_p$). The black dashed line here represents $\eta = 2.5$ mark, above which the shocks are capable of changing the initial injection spectra of particles (i.e. $p=3$) and hence the emission. The histogram is represented in the range of $\eta$ from 1.15 to 4 which corresponds to weakest to strongest shocks respectively. See Section \ref{Influence of diffusive shocks on particle spectra} for more details.}
    \label{fig:cmpr}
\end{figure*}

In Section \ref{sec:dynamics}, we have highlighted the fact that our simulated galaxy eventually becomes a weak shock system over time. Here, we will discuss whether these shocks can modify the particle spectra and hence the emission of it or not. In order to quantify this, we have analysed 1D-histograms of shock compression ratio ($\eta$) of the injected Lagrangian particles at three simulation times i.e. at 1.12 Myr, 1.49 Myr and 2.23 Myr. It is represented in Fig. \ref{fig:cmpr} within the range of 1.15 (weakest shocks) to 4 (strongest shocks) and normalized with respect to the maximum bin heights. We note here that a shock with compression ratio value of 2.5 or above can update the particle spectra to a value flatter than $p = 3$ which is the initial injection spectra. The point of $\eta = 2.5$ is highlighted by the black dashed line in Fig. \ref{fig:cmpr}. From figure, it is clear that at an initial stage of evolution of the jet, there exists significant number of particles that undergo shocks with $\eta \geq 2.5$ ($40\%$). Despite the number decreases with time, we still obtained $\sim 30\%$ particles with $\eta \geq 2.5$ at time 2.23 Myr. These are moderate to strongly shocked particles that show enhanced emission, effect of which has already been captured in our emission maps. 

On a similar note, the cooling processes also impart their signatures on the particle spectra by steepening it with time. For the synchrotron cooling of particles having $\gamma$ values $\geq 10^5$, the cooling time becomes $\leq 0.068$ Myr for an average magnetic field of 60 $\mu$G obtained for the jet. With a typical jet velocity similar to that of \textit{CaseB}, the particles can traverse a distance of $\leq 0.6$ kpc, before it starts dissipating to lower $\gamma$ values. The above estimation is a firm upper limit considering other cooling mechanisms in effect such as adiabatic and IC-CMB losses as well as the deceleration of the jet itself. However, we can find substantial patches of these particles (i.e. particles with $\gamma \geq 10^5$) beyond the estimated distance in Fig. \ref{fig:gamma_max}, dispersing throughout the cocoon. Here, the weak to moderate shocks have played a crucial role in pausing the drastic cooling of these non-thermal electrons that critically kept the structure active despite their aging with time \citep{Giri2022}.

In summary, diffusive shock acceleration plays a significant and continuous role in the overall evolution of galactic emission from jets.

\section{Equipartition approximation and Age estimation} \label{Equipartition approximation and Age estimation}
While measuring magnetic field strength in radio galaxies from synchrotron radiation, it is often assumed that the energy of the radiating electrons is in equipartition with the magnetic energy of the cocoon (see Eq. \ref{eq:equ}). The magnetic field estimated from this assumption is denoted as equipartition magnetic field strength ($B_{\rm eq}$) which is further used in estimating the age, luminosity, jet kinetic power of the source. However, in many radio galaxies, a significant deviation from this condition has been reported \citep{Croston2005,Croston2008,Hardcastle2010,Croston2011,mahatma2020}. As a result, the spectral age derived for these radio galaxies using $B_{\rm eq}$ yields a different result than their estimated dynamical age \citep{mahatma2020}. A wrong estimation of age further affects the estimation of other parameters that depends on it e.g. radio luminosity of the source \citep{Hodges-kluck2010}. So, it is crucial to understand the evolution of $B_{\rm eq}$ in these galaxies. 

\begin{figure}
    \centering
    \includegraphics[width=1\columnwidth]{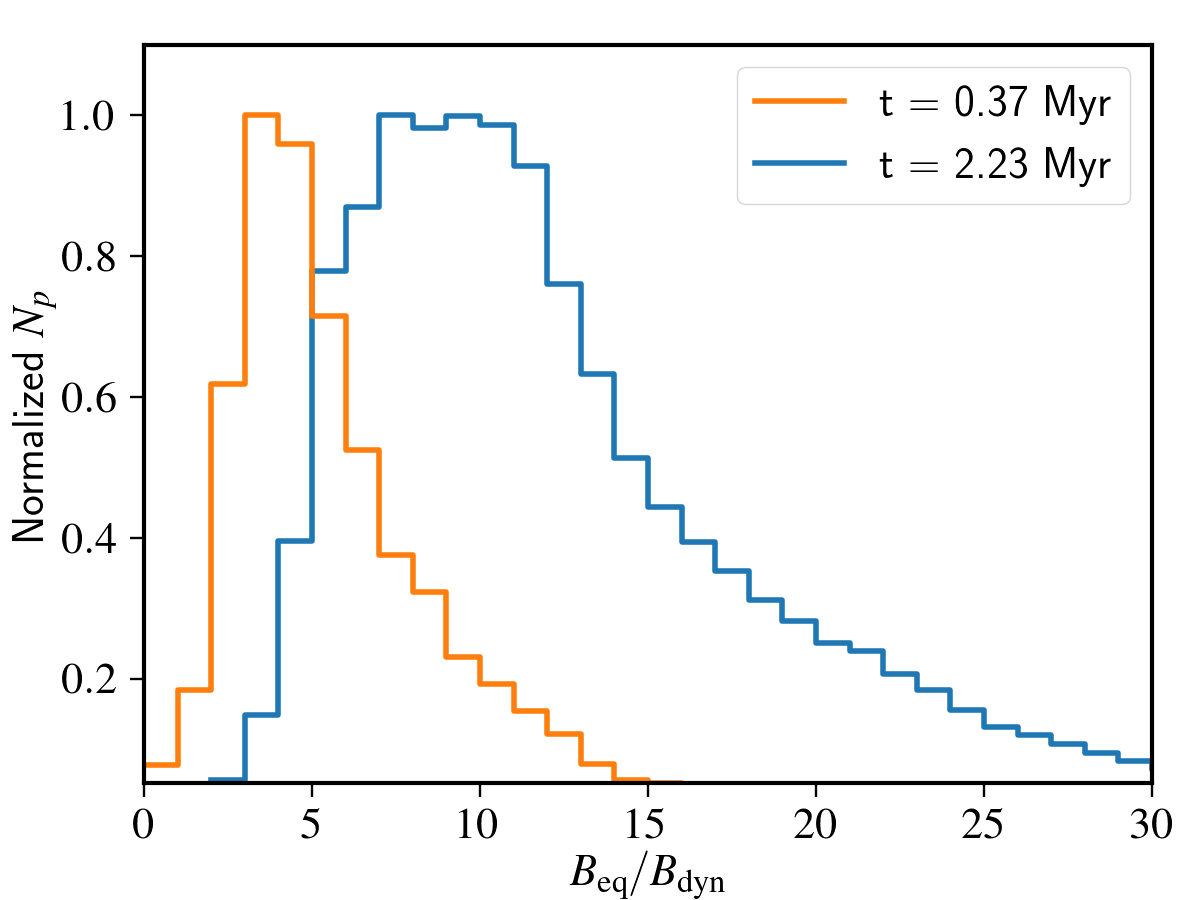}
    \caption{Histogram showing the time evolution of equipartition to dynamical magnetic field ratio ($B_{\rm eq}/B_{\rm dyn}$) for our synthetic radio structure. It shows a systematic shift of peak towards higher values of B-field ratio with time (t). The vertical axis represents here the normalized number of particles ($N_p$). This figure clearly demonstrates that our simulated structure at time 2.23 Myr deviate significantly from the equipartition condition.}
    \label{fig:histogram}
\end{figure}

For our reference case (i.e. \textit{CaseB}), we have computed $B_{\rm eq}$ (using particle spectra: Eq. \ref{eq:equ}) for all the particles injected into the domain and created a histogram of $\frac{B_{\rm eq}}{B_{\rm dyn}}$ at simulation time t $=$ 0.37 Myr and 2.23 Myr as shown in Fig. \ref{fig:histogram}. Here, $B_{\rm dyn}$ is the dynamical magnetic field evaluated at particle's position such that $B_{\rm dyn} = B_{\rm eq}$ represents true equipartition. At the initial stage (i.e.  at $\rm{t} = 0$), the histogram is expected to show a peaked distribution having $B_{\rm eq} = 0.18 B_{\rm dyn}$ (see Section \ref{Emission model}). With time, the distribution starts to diffuse and shifts towards higher values of the magnetic field ratio. At time $\rm{t} = 2.23$ Myr, the histogram peaks at a median value of 7 by showing a systematic shift of $\sim 2$ times the peak observed at $\rm{t} = 0.37$ Myr. The high magnetic field ratio tail of the histogram and its evolutionary trend with time suggest that the B-field ratio is most likely to shift towards higher values if it is allowed to evolve for a longer time, departing greatly from the equipartition state. The present deviation from equipartion of the simulated structure by a factor of 7 is not unusual as has been shown for radio galaxies by \citet{mahatma2020}.

We shall now demonstrate how age estimation is affected by our synthetic structure's deviation from equipartition criterion. The dynamical age of  
the formed structure is obtained as the age up to which the simulation has been conducted (t). The associate spectral age ($\rm{t}_{\rm spec}$) for an equipartition field strength of $B_{\rm eq}$ is evaluated as 
\begin{equation} \label{eq:spec_eq}
    {\rm t}_{\rm spec} = 50.3\frac{B_{\rm eq}^{1/2}}{B_{\rm eq}^2 + B_{\rm CMB}^2}((1+z)\nu_b)^{-1/2} \ \ \rm{Myr}
\end{equation}
Here, $B_{\rm eq}$ is estimated in nT, $z$ is the redshift of the galaxy adopted to be 0.0584 \citepalias{rubinur2017}, $B_{\rm CMB} = 0.318(1+z)^2$ nT is the magnetic field equivalent to cosmic microwave background radiation and $\nu_b$ is the break frequency in GHz \citep{Kardashev1962,Pacholczyk1970,Pandge2021}. Now at time t $=$ 2.23 Myr, the density-weighted average magnetic field ($B_{\rm dyn}$) is obtained as 60 $\mu$G (i.e. 6 $nT$) which corresponds to an equipartition field strength ($B_{\rm eq}$) of 42 nT. This has been calculated using the median value ($\sim 7$) of the $\frac{B_{\rm eq}}{B_{\rm dyn}}$ distribution at time t $=$ 2.23 Myr (see Fig. \ref{fig:histogram}). Using break frequency of the galaxy to be 11.5 GHz as adopted by \citetalias{rubinur2017}, we obtain ${\rm t}_{\rm spec} = 0.053$ Myr. This estimated spectral age for our synthetic radio structure is lower than the dynamical age (i.e. 2.23 Myr) as is typically found in observational studies. So, this is a result that has been generated from the galaxy's deviation from the equipartition state. 

The obtained $\rm{t}_{\rm spec}$ value of 0.053 Myr is still slightly lower than the spectral age estimated for 2MASX J1203 \citepalias{rubinur2017} which is 0.13 Myr. This additional deviation can be attributed to the assumptions that are opted when determining magnetic field strength from observable quantities. It's important to be aware of the uncertainties that come while estimating the B-field strength. The estimation is sensitive to the choices of parameters such as the spectral index value, extent of the source along the line of sight, lower cutoff frequency of the radio spectrum, volume filling factor, ratio of the relativistic proton to electron energy \footnote{\url{https://ned.ipac.caltech.edu/level5/Sept04/Govoni/Govoni3_2.html}\label{2}}. A different choices of them can result in a magnetic field strength ($B_{\rm eq}$) that is several times higher \textsuperscript{\ref{2}} than the predicted value provided in \citetalias{rubinur2017}, reducing further the value of spectral age (Eq. \ref{eq:spec_eq}) from 0.13 Myr.

\section{Summary}
\label{sec:summary}
We have performed high-resolution 3D MHD simulations of a precessing jet in a galactic environment to study the formation of an S-shaped winged galaxy using the hybrid Eulerian-Lagrangian framework of PLUTO code. Particularly, by adopting the parameter values suggested for the astronomical source 2MASX J1203 (having a prominent S-shape) in \citet{rubinur2017}, we have tried to reproduce its winged morphology through numerical means based on the jet precession model. The associated non-thermal emission of the formed structure due to synchrotron radiation is also modelled considering the role of adiabatic and radiative cooling and the diffusive shock acceleration on the particle spectral evolution. Such a hybrid approach would provide detailed insight into the radiative processes that are occurring throughout the galaxy which is a novel approach.
Our study focuses on a particular winged source which has a typical extension of $< 4$ kpc and is believed to be originating from a Seyfert type galaxy with a probable binary AGN system at the core.

The key results from the present work are summarized as follows:
\begin{itemize}
    \item \textbf{Dynamical evolution:} We show here that a precessing jet can produce a distinct S-shaped morphology if the jet evolves for a sufficient time. This is required to limit the contribution of the bow shocked region in the emission map, as the strength of the bow shock weakens over time. These jetted morphology eventually settles in an equal distribution of the cocoon's thermal and magnetic energy over time, with a significant deviation from equipartition between the radiating electrons energy and the magnetic energy.
    
    The simulated morphology is sensitive to the parameter choices of the jet. The size and geometry of the resulting structure are largely influenced by the increased interaction between the jet and the ambient medium, as well as the jet's self-deceleration due to MHD instabilities. As a result, the kinematic model of a precessing jet \citep[e.g.,][]{hjellming1981} needs to be corrected for these deceleration effects in order to properly predict the precession parameters. Numerical simulations in this context become an efficient way of constraining the jet and ambient medium parameters where the effects of deceleration are obtained self consistently.
    \item \textbf{Spectral properties and polarization:} The fact that the observed winged morphology of 2MASX J1203 can form based on a precessing jet's evolution in a galactic environment is clearly supported by our synthetic emission maps. A large scale analogy of this structure has recently been reported by \citet{Bruni2021} who have alluded its formation mechanism to a precessing jet as well. These sources are a strong contender for hosting a binary super massive black hole system at their core which would necessitate further high resolution observations to detect them, hence resolving their formation scenarios.  
    
    The spectral index ($\alpha$) map of our simulated structure shows that the hotspots possess flat $\alpha$-values. The jet ridge lines possess moderately steep values of $\alpha$, compared to the jet side and leading edges. Additionally, we have synthetically produced the polarisation map of the radio structure that would serve as a template for future observational proposals. In particular, we have seen magnetic field lines aligning with the jet flow which is formed due to the jet ambient medium interaction \citep{Laing1981,Roberts2008}. The polarisation fraction of the jet is high, reaching up to 76\% in side and leading edges of the jet, indicating the presence of a highly ordered B-field configuration caused by moderate to strong shocks. We have further demonstrated how important these diffusive shocks are in keeping the structure active throughout its evolution by pausing the drastic cooling of radiating electrons as well as re-energizing them.
    \item \textbf{Equipartition approximation and age estimation:} We have established here that the produced morphology's equipartition condition (between the radiating electrons energy and the magnetic energy) changes with time, governed by the combined effect of radiative losses from local magnetic fields and diffusive shock acceleration. Starting with an initial under-equipartition state ($B_{\rm eq} < B_{\rm dyn}$), the radio structure slowly evolves to an above-equipartition stage ($B_{\rm eq} > B_{\rm dyn}$) by showing a systematic shift in $\frac{B_{\rm eq}}{B_{\rm dyn}}$ values. This has further implications in the age estimation of such sources. The spectral age evaluated for the formed morphology ($0.053$ Myr) is less than the dynamical age ($2.23$ Myr), as is typically found in observational studies.
\end{itemize}
The present study showcases the fact that the jet precession model is capable of producing the morphology typically observed in winged sources, whose formation scenario is long being debated \citep[][]{Rottmann2002,Gopal-Krishna2012,Giri2022}. However, further studies are required in order to confirm this model's universality by extending it to larger scaled winged sources, as is typically found in observations \citep[see][]{Capetti2002,Bruni2021}. We also note here that our simulations are not designed to resolve the origin of jet precession that is causing the winged structure to form. So further studies can be taken up to better connect the small scale simulations (including physics of unstable accretion disk or binary BH in orbit or even inhomogeneous accretion flow to BH \citep[e.g.,][]{Liska2018,Lalakos2022}) with the higher scaled peculiar morphologies of these sources.  
\section*{Acknowledgement}
The authors would like to thank the referee for providing valuable suggestions and comments, which helped in improving the manuscript significantly.
BV would like to thank The Council of Scientific \& Industrial Research (CSIR) [03(1435)/18/EMR-II] and Max Planck Partner Group Award. GG is supported by the Prime Minister's Research Fellowship. Numerical simulations presented in this work are performed using facilities at Indian Institute of Technology Indore and MPG Super-computing Cobra cluster (\url{https://www.mpcdf.mpg.de/services/supercomputing/cobra}). 
\section*{Data Availability}
The source files of our simulations and the generated data used in this work will be shared on reasonable request to the corresponding authors.



\bibliographystyle{mnras}
\bibliography{references} 

\begin{thebibliography}{}
\makeatletter
\relax
\def\mn@urlcharsother{\let\do\@makeother \do\$\do\&\do\#\do\^\do\_\do\%\do\~}
\def\mn@doi{\begingroup\mn@urlcharsother \@ifnextchar [ {\mn@doi@}
  {\mn@doi@[]}}
\def\mn@doi@[#1]#2{\def\@tempa{#1}\ifx\@tempa\@empty \href
  {http://dx.doi.org/#2} {doi:#2}\else \href {http://dx.doi.org/#2} {#1}\fi
  \endgroup}
\def\mn@eprint#1#2{\mn@eprint@#1:#2::\@nil}
\def\mn@eprint@arXiv#1{\href {http://arxiv.org/abs/#1} {{\tt arXiv:#1}}}
\def\mn@eprint@dblp#1{\href {http://dblp.uni-trier.de/rec/bibtex/#1.xml}
  {dblp:#1}}
\def\mn@eprint@#1:#2:#3:#4\@nil{\def\@tempa {#1}\def\@tempb {#2}\def\@tempc
  {#3}\ifx \@tempc \@empty \let \@tempc \@tempb \let \@tempb \@tempa \fi \ifx
  \@tempb \@empty \def\@tempb {arXiv}\fi \@ifundefined
  {mn@eprint@\@tempb}{\@tempb:\@tempc}{\expandafter \expandafter \csname
  mn@eprint@\@tempb\endcsname \expandafter{\@tempc}}}

\bibitem[\protect\citeauthoryear{{Acharya}, {Borse}  \& {Vaidya}}{{Acharya}
  et~al.}{2021}]{Acharya2021}
{Acharya} S.,  {Borse} N.~S.,   {Vaidya} B.,  2021, \mn@doi [\mnras]
  {10.1093/mnras/stab1775}, \href
  {https://ui.adsabs.harvard.edu/abs/2021MNRAS.506.1862A} {506, 1862}

\bibitem[\protect\citeauthoryear{{Bassani}, {Venturi}, {Molina}, {Malizia},
  {Dallacasa}, {Panessa}, {Bazzano}  \& {Ubertini}}{{Bassani}
  et~al.}{2016}]{Bassani2016}
{Bassani} L.,  {Venturi} T.,  {Molina} M.,  {Malizia} A.,  {Dallacasa} D.,
  {Panessa} F.,  {Bazzano} A.,   {Ubertini} P.,  2016, \mn@doi [\mnras]
  {10.1093/mnras/stw1468}, \href
  {https://ui.adsabs.harvard.edu/abs/2016MNRAS.461.3165B} {461, 3165}

\bibitem[\protect\citeauthoryear{{Baty} \& {Keppens}}{{Baty} \&
  {Keppens}}{2002}]{Baty2002}
{Baty} H.,  {Keppens} R.,  2002, \mn@doi [\apj] {10.1086/343893}, \href
  {https://ui.adsabs.harvard.edu/abs/2002ApJ...580..800B} {580, 800}

\bibitem[\protect\citeauthoryear{{Begelman}, {Blandford}  \& {Rees}}{{Begelman}
  et~al.}{1980}]{begelman1980}
{Begelman} M.~C.,  {Blandford} R.~D.,   {Rees} M.~J.,  1980, \mn@doi [\nat]
  {10.1038/287307a0}, \href
  {https://ui.adsabs.harvard.edu/abs/1980Natur.287..307B} {287, 307}

\bibitem[\protect\citeauthoryear{{Bodo}, {Mamatsashvili}, {Rossi}  \&
  {Mignone}}{{Bodo} et~al.}{2013}]{Bodo2013}
{Bodo} G.,  {Mamatsashvili} G.,  {Rossi} P.,   {Mignone} A.,  2013, \mn@doi
  [\mnras] {10.1093/mnras/stt1225}, \href
  {https://ui.adsabs.harvard.edu/abs/2013MNRAS.434.3030B} {434, 3030}

\bibitem[\protect\citeauthoryear{{Borse}, {Acharya}, {Vaidya}, {Mukherjee},
  {Bodo}, {Rossi}  \& {Mignone}}{{Borse} et~al.}{2021}]{Borse2021}
{Borse} N.,  {Acharya} S.,  {Vaidya} B.,  {Mukherjee} D.,  {Bodo} G.,  {Rossi}
  P.,   {Mignone} A.,  2021, \mn@doi [\aap] {10.1051/0004-6361/202140440},
  \href {https://ui.adsabs.harvard.edu/abs/2021A&A...649A.150B} {649, A150}

\bibitem[\protect\citeauthoryear{{Bruni} et~al.,}{{Bruni}
  et~al.}{2021}]{Bruni2021}
{Bruni} G.,  et~al., 2021, \mn@doi [\mnras] {10.1093/mnras/stab623}, \href
  {https://ui.adsabs.harvard.edu/abs/2021MNRAS.503.4681B} {503, 4681}

\bibitem[\protect\citeauthoryear{{Capetti}, {Zamfir}, {Rossi}, {Bodo}, {Zanni}
  \& {Massaglia}}{{Capetti} et~al.}{2002}]{Capetti2002}
{Capetti} A.,  {Zamfir} S.,  {Rossi} P.,  {Bodo} G.,  {Zanni} C.,   {Massaglia}
  S.,  2002, \mn@doi [\aap] {10.1051/0004-6361:20021070}, \href
  {https://ui.adsabs.harvard.edu/abs/2002A&A...394...39C} {394, 39}

\bibitem[\protect\citeauthoryear{{Cavaliere} \& {Fusco-Femiano}}{{Cavaliere} \&
  {Fusco-Femiano}}{1978}]{cavaliere1978}
{Cavaliere} A.,  {Fusco-Femiano} R.,  1978, \aap, \href
  {https://ui.adsabs.harvard.edu/abs/1978A&A....70..677C} {70, 677}

\bibitem[\protect\citeauthoryear{{Cotton} et~al.,}{{Cotton}
  et~al.}{2020}]{Cotton2020}
{Cotton} W.~D.,  et~al., 2020, \mn@doi [\mnras] {10.1093/mnras/staa1240}, \href
  {https://ui.adsabs.harvard.edu/abs/2020MNRAS.495.1271C} {495, 1271}

\bibitem[\protect\citeauthoryear{{Croston}, {Hardcastle}, {Harris}, {Belsole},
  {Birkinshaw}  \& {Worrall}}{{Croston} et~al.}{2005}]{Croston2005}
{Croston} J.~H.,  {Hardcastle} M.~J.,  {Harris} D.~E.,  {Belsole} E.,
  {Birkinshaw} M.,   {Worrall} D.~M.,  2005, \mn@doi [\apj] {10.1086/430170},
  \href {https://ui.adsabs.harvard.edu/abs/2005ApJ...626..733C} {626, 733}

\bibitem[\protect\citeauthoryear{{Croston}, {Hardcastle}, {Birkinshaw},
  {Worrall}  \& {Laing}}{{Croston} et~al.}{2008}]{Croston2008}
{Croston} J.~H.,  {Hardcastle} M.~J.,  {Birkinshaw} M.,  {Worrall} D.~M.,
  {Laing} R.~A.,  2008, \mn@doi [\mnras] {10.1111/j.1365-2966.2008.13162.x},
  \href {https://ui.adsabs.harvard.edu/abs/2008MNRAS.386.1709C} {386, 1709}

\bibitem[\protect\citeauthoryear{{Croston}, {Hardcastle}, {Mingo}, {Evans},
  {Dicken}, {Morganti}  \& {Tadhunter}}{{Croston} et~al.}{2011}]{Croston2011}
{Croston} J.~H.,  {Hardcastle} M.~J.,  {Mingo} B.,  {Evans} D.~A.,  {Dicken}
  D.,  {Morganti} R.,   {Tadhunter} C.~N.,  2011, \mn@doi [\apjl]
  {10.1088/2041-8205/734/2/L28}, \href
  {https://ui.adsabs.harvard.edu/abs/2011ApJ...734L..28C} {734, L28}

\bibitem[\protect\citeauthoryear{{Dabhade} et~al.,}{{Dabhade}
  et~al.}{2020}]{Dabhade2020}
{Dabhade} P.,  et~al., 2020, \mn@doi [\aap] {10.1051/0004-6361/202038344},
  \href {https://ui.adsabs.harvard.edu/abs/2020A&A...642A.153D} {642, A153}

\bibitem[\protect\citeauthoryear{{Dedner}, {Kemm}, {Kr{\"o}ner}, {Munz},
  {Schnitzer}  \& {Wesenberg}}{{Dedner} et~al.}{2002}]{Dedner2002}
{Dedner} A.,  {Kemm} F.,  {Kr{\"o}ner} D.,  {Munz} C.~D.,  {Schnitzer} T.,
  {Wesenberg} M.,  2002, \mn@doi [Journal of Computational Physics]
  {10.1006/jcph.2001.6961}, \href
  {https://ui.adsabs.harvard.edu/abs/2002JCoPh.175..645D} {175, 645}

\bibitem[\protect\citeauthoryear{{Dennett-Thorpe}, {Scheuer}, {Laing},
  {Bridle}, {Pooley}  \& {Reich}}{{Dennett-Thorpe}
  et~al.}{2002}]{Dennett-Thorpe2002}
{Dennett-Thorpe} J.,  {Scheuer} P.~A.~G.,  {Laing} R.~A.,  {Bridle} A.~H.,
  {Pooley} G.~G.,   {Reich} W.,  2002, \mn@doi [\mnras]
  {10.1046/j.1365-8711.2002.05106.x}, \href
  {https://ui.adsabs.harvard.edu/abs/2002MNRAS.330..609D} {330, 609}

\bibitem[\protect\citeauthoryear{Ekers}{Ekers}{1982}]{ekers1982}
Ekers R.,  1982, in Symposium-International Astronomical Union. pp 465--474

\bibitem[\protect\citeauthoryear{{Ekers}, {Fanti}, {Lari}  \& {Parma}}{{Ekers}
  et~al.}{1978}]{Ekers1978}
{Ekers} R.~D.,  {Fanti} R.,  {Lari} C.,   {Parma} P.,  1978, \mn@doi [\nat]
  {10.1038/276588a0}, \href
  {https://ui.adsabs.harvard.edu/abs/1978Natur.276..588E} {276, 588}

\bibitem[\protect\citeauthoryear{{Falceta-Gon{\c{c}}alves}, {Caproni},
  {Abraham}, {Teixeira}  \& {de Gouveia Dal Pino}}{{Falceta-Gon{\c{c}}alves}
  et~al.}{2010}]{Falceta2010}
{Falceta-Gon{\c{c}}alves} D.,  {Caproni} A.,  {Abraham} Z.,  {Teixeira} D.~M.,
   {de Gouveia Dal Pino} E.~M.,  2010, \mn@doi [\apjl]
  {10.1088/2041-8205/713/1/L74}, \href
  {https://ui.adsabs.harvard.edu/abs/2010ApJ...713L..74F} {713, L74}

\bibitem[\protect\citeauthoryear{{Fan}, {Liu}, {Wang}, {Fryer}  \& {Li}}{{Fan}
  et~al.}{2008}]{Fan2008}
{Fan} Z.-H.,  {Liu} S.,  {Wang} J.-M.,  {Fryer} C.~L.,   {Li} H.,  2008,
  \mn@doi [\apjl] {10.1086/528372}, \href
  {https://ui.adsabs.harvard.edu/abs/2008ApJ...673L.139F} {673, L139}

\bibitem[\protect\citeauthoryear{{Fu}, {Myers}, {Djorgovski}  \& {Yan}}{{Fu}
  et~al.}{2011}]{Fu2011}
{Fu} H.,  {Myers} A.~D.,  {Djorgovski} S.~G.,   {Yan} L.,  2011, \mn@doi [\apj]
  {10.1088/0004-637X/733/2/103}, \href
  {https://ui.adsabs.harvard.edu/abs/2011ApJ...733..103F} {733, 103}

\bibitem[\protect\citeauthoryear{{Fu}, {Yan}, {Myers}, {Stockton},
  {Djorgovski}, {Aldering}  \& {Rich}}{{Fu} et~al.}{2012}]{Fu2012}
{Fu} H.,  {Yan} L.,  {Myers} A.~D.,  {Stockton} A.,  {Djorgovski} S.~G.,
  {Aldering} G.,   {Rich} J.~A.,  2012, \mn@doi [\apj]
  {10.1088/0004-637X/745/1/67}, \href
  {https://ui.adsabs.harvard.edu/abs/2012ApJ...745...67F} {745, 67}

\bibitem[\protect\citeauthoryear{{Giri}, {Vaidya}, {Rossi}, {Bodo}, {Mukherjee}
   \& {Mignone}}{{Giri} et~al.}{2022}]{Giri2022}
{Giri} G.,  {Vaidya} B.,  {Rossi} P.,  {Bodo} G.,  {Mukherjee} D.,   {Mignone}
  A.,  2022, \mn@doi [\aap] {10.1051/0004-6361/202142546}, \href
  {https://ui.adsabs.harvard.edu/abs/2022A&A...662A...5G} {662, A5}

\bibitem[\protect\citeauthoryear{{Gopal-Krishna}, {Biermann}  \&
  {Wiita}}{{Gopal-Krishna} et~al.}{2003}]{Gopal-Krishna2003}
{Gopal-Krishna} {Biermann} P.~L.,   {Wiita} P.~J.,  2003, \mn@doi [\apjl]
  {10.1086/378766}, \href
  {https://ui.adsabs.harvard.edu/abs/2003ApJ...594L.103G} {594, L103}

\bibitem[\protect\citeauthoryear{{Gopal-Krishna}, {Biermann}, {Gergely}  \&
  {Wiita}}{{Gopal-Krishna} et~al.}{2012}]{Gopal-Krishna2012}
{Gopal-Krishna} {Biermann} P.~L.,  {Gergely} L.~{\'A}.,   {Wiita} P.~J.,  2012,
  \mn@doi [Research in Astronomy and Astrophysics]
  {10.1088/1674-4527/12/2/002}, \href
  {https://ui.adsabs.harvard.edu/abs/2012RAA....12..127G} {12, 127}

\bibitem[\protect\citeauthoryear{{Gower}, {Gregory}, {Unruh}  \&
  {Hutchings}}{{Gower} et~al.}{1982}]{Gower1982}
{Gower} A.~C.,  {Gregory} P.~C.,  {Unruh} W.~G.,   {Hutchings} J.~B.,  1982,
  \mn@doi [\apj] {10.1086/160442}, \href
  {https://ui.adsabs.harvard.edu/abs/1982ApJ...262..478G} {262, 478}

\bibitem[\protect\citeauthoryear{{Hardcastle}}{{Hardcastle}}{2010}]{Hardcastle2010}
{Hardcastle} M.~J.,  2010, \mn@doi [\mnras] {10.1111/j.1365-2966.2010.16668.x},
  \href {https://ui.adsabs.harvard.edu/abs/2010MNRAS.405.2810H} {405, 2810}

\bibitem[\protect\citeauthoryear{{Hardcastle} \& {Krause}}{{Hardcastle} \&
  {Krause}}{2014}]{hardcastle2014}
{Hardcastle} M.~J.,  {Krause} M.~G.~H.,  2014, \mn@doi [\mnras]
  {10.1093/mnras/stu1229}, \href
  {https://ui.adsabs.harvard.edu/abs/2014MNRAS.443.1482H} {443, 1482}

\bibitem[\protect\citeauthoryear{{Hardcastle}, {Birkinshaw}, {Cameron},
  {Harris}, {Looney}  \& {Worrall}}{{Hardcastle} et~al.}{2002}]{Hardcastle2002}
{Hardcastle} M.~J.,  {Birkinshaw} M.,  {Cameron} R.~A.,  {Harris} D.~E.,
  {Looney} L.~W.,   {Worrall} D.~M.,  2002, \mn@doi [\apj] {10.1086/344409},
  \href {https://ui.adsabs.harvard.edu/abs/2002ApJ...581..948H} {581, 948}

\bibitem[\protect\citeauthoryear{{Hardcastle} et~al.,}{{Hardcastle}
  et~al.}{2019}]{Hardcastle2019}
{Hardcastle} M.~J.,  et~al., 2019, \mn@doi [\mnras] {10.1093/mnras/stz1910},
  \href {https://ui.adsabs.harvard.edu/abs/2019MNRAS.488.3416H} {488, 3416}

\bibitem[\protect\citeauthoryear{{Hardee}, {Hughes}, {Rosen}  \&
  {Gomez}}{{Hardee} et~al.}{2001}]{Hardee2001}
{Hardee} P.~E.,  {Hughes} P.~A.,  {Rosen} A.,   {Gomez} E.~A.,  2001, \mn@doi
  [\apj] {10.1086/321525}, \href
  {https://ui.adsabs.harvard.edu/abs/2001ApJ...555..744H} {555, 744}

\bibitem[\protect\citeauthoryear{{Hjellming} \& {Johnston}}{{Hjellming} \&
  {Johnston}}{1981}]{hjellming1981}
{Hjellming} R.~M.,  {Johnston} K.~J.,  1981, \mn@doi [\apjl] {10.1086/183571},
  \href {https://ui.adsabs.harvard.edu/abs/1981ApJ...246L.141H} {246, L141}

\bibitem[\protect\citeauthoryear{{Hodges-Kluck} \& {Reynolds}}{{Hodges-Kluck}
  \& {Reynolds}}{2012}]{Hodges-kluck2012}
{Hodges-Kluck} E.~J.,  {Reynolds} C.~S.,  2012, \mn@doi [\apj]
  {10.1088/0004-637X/746/2/167}, \href
  {https://ui.adsabs.harvard.edu/abs/2012ApJ...746..167H} {746, 167}

\bibitem[\protect\citeauthoryear{{Hodges-Kluck}, {Reynolds}, {Miller}  \&
  {Cheung}}{{Hodges-Kluck} et~al.}{2010}]{Hodges-kluck2010}
{Hodges-Kluck} E.~J.,  {Reynolds} C.~S.,  {Miller} M.~C.,   {Cheung} C.~C.,
  2010, \mn@doi [\apjl] {10.1088/2041-8205/717/1/L37}, \href
  {https://ui.adsabs.harvard.edu/abs/2010ApJ...717L..37H} {717, L37}

\bibitem[\protect\citeauthoryear{{Horton}, {Krause}  \& {Hardcastle}}{{Horton}
  et~al.}{2020}]{horton2020}
{Horton} M.~A.,  {Krause} M. G.~H.,   {Hardcastle} M.~J.,  2020, \mn@doi
  [\mnras] {10.1093/mnras/staa3020}, \href
  {https://ui.adsabs.harvard.edu/abs/2020MNRAS.499.5765H} {499, 5765}

\bibitem[\protect\citeauthoryear{{Kardashev}}{{Kardashev}}{1962}]{Kardashev1962}
{Kardashev} N.~S.,  1962, \sovast, \href
  {https://ui.adsabs.harvard.edu/abs/1962SvA.....6..317K} {6, 317}

\bibitem[\protect\citeauthoryear{{Kellermann}, {Sramek}, {Schmidt}, {Shaffer}
  \& {Green}}{{Kellermann} et~al.}{1989}]{kellermann1989}
{Kellermann} K.~I.,  {Sramek} R.,  {Schmidt} M.,  {Shaffer} D.~B.,   {Green}
  R.,  1989, \mn@doi [\aj] {10.1086/115207}, \href
  {https://ui.adsabs.harvard.edu/abs/1989AJ.....98.1195K} {98, 1195}

\bibitem[\protect\citeauthoryear{{Kharb}, {Das}, {Paragi}, {Subramanian}  \&
  {Chitta}}{{Kharb} et~al.}{2015}]{Kharb2015}
{Kharb} P.,  {Das} M.,  {Paragi} Z.,  {Subramanian} S.,   {Chitta} L.~P.,
  2015, \mn@doi [\apj] {10.1088/0004-637X/799/2/161}, \href
  {https://ui.adsabs.harvard.edu/abs/2015ApJ...799..161K} {799, 161}

\bibitem[\protect\citeauthoryear{{Kharb}, {Vaddi}, {Sebastian}, {Subramanian},
  {Das}  \& {Paragi}}{{Kharb} et~al.}{2019}]{Kharb2019}
{Kharb} P.,  {Vaddi} S.,  {Sebastian} B.,  {Subramanian} S.,  {Das} M.,
  {Paragi} Z.,  2019, \mn@doi [\apj] {10.3847/1538-4357/aafad7}, \href
  {https://ui.adsabs.harvard.edu/abs/2019ApJ...871..249K} {871, 249}

\bibitem[\protect\citeauthoryear{{Krause} et~al.,}{{Krause}
  et~al.}{2019}]{Krause2019}
{Krause} M. G.~H.,  et~al., 2019, \mn@doi [\mnras] {10.1093/mnras/sty2558},
  \href {https://ui.adsabs.harvard.edu/abs/2019MNRAS.482..240K} {482, 240}

\bibitem[\protect\citeauthoryear{{Kurosawa} \& {Proga}}{{Kurosawa} \&
  {Proga}}{2008}]{Kurosawa2008}
{Kurosawa} R.,  {Proga} D.,  2008, \mn@doi [\apj] {10.1086/524870}, \href
  {https://ui.adsabs.harvard.edu/abs/2008ApJ...674...97K} {674, 97}

\bibitem[\protect\citeauthoryear{{Laing}}{{Laing}}{1981}]{Laing1981}
{Laing} R.~A.,  1981, \mn@doi [\apj] {10.1086/159132}, \href
  {https://ui.adsabs.harvard.edu/abs/1981ApJ...248...87L} {248, 87}

\bibitem[\protect\citeauthoryear{{Lalakos} et~al.,}{{Lalakos}
  et~al.}{2022}]{Lalakos2022}
{Lalakos} A.,  et~al., 2022, arXiv e-prints, \href
  {https://ui.adsabs.harvard.edu/abs/2022arXiv220208281L} {p. arXiv:2202.08281}

\bibitem[\protect\citeauthoryear{{Leahy} \& {Williams}}{{Leahy} \&
  {Williams}}{1984}]{Leahy1984}
{Leahy} J.~P.,  {Williams} A.~G.,  1984, \mn@doi [\mnras]
  {10.1093/mnras/210.4.929}, \href
  {https://ui.adsabs.harvard.edu/abs/1984MNRAS.210..929L} {210, 929}

\bibitem[\protect\citeauthoryear{{Lind}, {Payne}, {Meier}  \&
  {Blandford}}{{Lind} et~al.}{1989}]{Lind1989}
{Lind} K.~R.,  {Payne} D.~G.,  {Meier} D.~L.,   {Blandford} R.~D.,  1989,
  \mn@doi [\apj] {10.1086/167779}, \href
  {https://ui.adsabs.harvard.edu/abs/1989ApJ...344...89L} {344, 89}

\bibitem[\protect\citeauthoryear{{Liska}, {Hesp}, {Tchekhovskoy}, {Ingram},
  {van der Klis}  \& {Markoff}}{{Liska} et~al.}{2018}]{Liska2018}
{Liska} M.,  {Hesp} C.,  {Tchekhovskoy} A.,  {Ingram} A.,  {van der Klis} M.,
  {Markoff} S.,  2018, \mn@doi [\mnras] {10.1093/mnrasl/slx174}, \href
  {https://ui.adsabs.harvard.edu/abs/2018MNRAS.474L..81L} {474, L81}

\bibitem[\protect\citeauthoryear{{Mahatma}, {Hardcastle}, {Croston}, {Harwood},
  {Ineson}  \& {Moldon}}{{Mahatma} et~al.}{2020}]{mahatma2020}
{Mahatma} V.~H.,  {Hardcastle} M.~J.,  {Croston} J.~H.,  {Harwood} J.,
  {Ineson} J.,   {Moldon} J.,  2020, \mn@doi [\mnras] {10.1093/mnras/stz3396},
  \href {https://ui.adsabs.harvard.edu/abs/2020MNRAS.491.5015M} {491, 5015}

\bibitem[\protect\citeauthoryear{{Merloni} \& {Heinz}}{{Merloni} \&
  {Heinz}}{2007}]{Merloni2007}
{Merloni} A.,  {Heinz} S.,  2007, \mn@doi [\mnras]
  {10.1111/j.1365-2966.2007.12253.x}, \href
  {https://ui.adsabs.harvard.edu/abs/2007MNRAS.381..589M} {381, 589}

\bibitem[\protect\citeauthoryear{{Mignone} \& {Bodo}}{{Mignone} \&
  {Bodo}}{2006}]{Mignone2006}
{Mignone} A.,  {Bodo} G.,  2006, \mn@doi [\mnras]
  {10.1111/j.1365-2966.2006.10162.x}, \href
  {https://ui.adsabs.harvard.edu/abs/2006MNRAS.368.1040M} {368, 1040}

\bibitem[\protect\citeauthoryear{{Mignone}, {Bodo}, {Massaglia}, {Matsakos},
  {Tesileanu}, {Zanni}  \& {Ferrari}}{{Mignone} et~al.}{2007}]{mignone2007}
{Mignone} A.,  {Bodo} G.,  {Massaglia} S.,  {Matsakos} T.,  {Tesileanu} O.,
  {Zanni} C.,   {Ferrari} A.,  2007, \mn@doi [\apjs] {10.1086/513316}, \href
  {https://ui.adsabs.harvard.edu/abs/2007ApJS..170..228M} {170, 228}

\bibitem[\protect\citeauthoryear{{Monceau-Baroux}, {Porth}, {Meliani}  \&
  {Keppens}}{{Monceau-Baroux} et~al.}{2014}]{Monceau2014}
{Monceau-Baroux} R.,  {Porth} O.,  {Meliani} Z.,   {Keppens} R.,  2014, \mn@doi
  [\aap] {10.1051/0004-6361/201322682}, \href
  {https://ui.adsabs.harvard.edu/abs/2014A&A...561A..30M} {561, A30}

\bibitem[\protect\citeauthoryear{{Monceau-Baroux}, {Porth}, {Meliani}  \&
  {Keppens}}{{Monceau-Baroux} et~al.}{2015}]{Monceau2015}
{Monceau-Baroux} R.,  {Porth} O.,  {Meliani} Z.,   {Keppens} R.,  2015, \mn@doi
  [\aap] {10.1051/0004-6361/201425015}, \href
  {https://ui.adsabs.harvard.edu/abs/2015A&A...574A.143M} {574, A143}

\bibitem[\protect\citeauthoryear{{Mukherjee}, {Bodo}, {Mignone}, {Rossi}  \&
  {Vaidya}}{{Mukherjee} et~al.}{2020}]{Mukherjee2020}
{Mukherjee} D.,  {Bodo} G.,  {Mignone} A.,  {Rossi} P.,   {Vaidya} B.,  2020,
  \mn@doi [\mnras] {10.1093/mnras/staa2934}, \href
  {https://ui.adsabs.harvard.edu/abs/2020MNRAS.499..681M} {499, 681}

\bibitem[\protect\citeauthoryear{{Mukherjee}, {Bodo}, {Rossi}, {Mignone}  \&
  {Vaidya}}{{Mukherjee} et~al.}{2021}]{Mukherjee2021}
{Mukherjee} D.,  {Bodo} G.,  {Rossi} P.,  {Mignone} A.,   {Vaidya} B.,  2021,
  arXiv e-prints, \href {https://ui.adsabs.harvard.edu/abs/2021arXiv210502836M}
  {p. arXiv:2105.02836}

\bibitem[\protect\citeauthoryear{{M{\"u}ller} et~al.,}{{M{\"u}ller}
  et~al.}{2021}]{Muller2021}
{M{\"u}ller} A.,  et~al., 2021, \mn@doi [\mnras] {10.1093/mnras/stab2928},
  \href {https://ui.adsabs.harvard.edu/abs/2021MNRAS.508.5326M} {508, 5326}

\bibitem[\protect\citeauthoryear{{Murgia}, {Parma}, {de Ruiter}, {Bondi},
  {Ekers}, {Fanti}  \& {Fomalont}}{{Murgia} et~al.}{2001}]{Murgia2001}
{Murgia} M.,  {Parma} P.,  {de Ruiter} H.~R.,  {Bondi} M.,  {Ekers} R.~D.,
  {Fanti} R.,   {Fomalont} E.~B.,  2001, \mn@doi [\aap]
  {10.1051/0004-6361:20011436}, \href
  {https://ui.adsabs.harvard.edu/abs/2001A&A...380..102M} {380, 102}

\bibitem[\protect\citeauthoryear{{Nawaz}, {Bicknell}, {Wagner}, {Sutherland}
  \& {McNamara}}{{Nawaz} et~al.}{2016}]{Nawaz2016}
{Nawaz} M.~A.,  {Bicknell} G.~V.,  {Wagner} A.~Y.,  {Sutherland} R.~S.,
  {McNamara} B.~R.,  2016, \mn@doi [\mnras] {10.1093/mnras/stw330}, \href
  {https://ui.adsabs.harvard.edu/abs/2016MNRAS.458..802N} {458, 802}

\bibitem[\protect\citeauthoryear{{O'Neill}, {Jones}, {Nolting}  \&
  {Mendygral}}{{O'Neill} et~al.}{2019}]{Oneill2019}
{O'Neill} B.~J.,  {Jones} T.~W.,  {Nolting} C.,   {Mendygral} P.~J.,  2019,
  \mn@doi [\apj] {10.3847/1538-4357/ab4efa}, \href
  {https://ui.adsabs.harvard.edu/abs/2019ApJ...887...26O} {887, 26}

\bibitem[\protect\citeauthoryear{{Pacholczyk}}{{Pacholczyk}}{1970}]{Pacholczyk1970}
{Pacholczyk} A.~G.,  1970, {Radio astrophysics. Nonthermal processes in
  galactic and extragalactic sources}

\bibitem[\protect\citeauthoryear{{Pandge}, {Kale}, {Dabhade}, {Mahato}  \&
  {Raychaudhury}}{{Pandge} et~al.}{2021}]{Pandge2021}
{Pandge} M.~B.,  {Kale} R.,  {Dabhade} P.,  {Mahato} M.,   {Raychaudhury} S.,
  2021, \mn@doi [\mnras] {10.1093/mnras/stab2945}, \href
  {https://ui.adsabs.harvard.edu/abs/2021MNRAS.tmp.2712P} {}

\bibitem[\protect\citeauthoryear{{Panferov}}{{Panferov}}{2014}]{Panferov2014}
{Panferov} A.,  2014, \mn@doi [\aap] {10.1051/0004-6361/201322456}, \href
  {https://ui.adsabs.harvard.edu/abs/2014A&A...562A.130P} {562, A130}

\bibitem[\protect\citeauthoryear{{Pringle}}{{Pringle}}{1996}]{Pringle1996}
{Pringle} J.~E.,  1996, \mn@doi [\mnras] {10.1093/mnras/281.1.357}, \href
  {https://ui.adsabs.harvard.edu/abs/1996MNRAS.281..357P} {281, 357}

\bibitem[\protect\citeauthoryear{{Querejeta}, {Eliche-Moral}, {Tapia},
  {Borlaff}, {Rodr{\'\i}guez-P{\'e}rez}, {Zamorano}  \& {Gallego}}{{Querejeta}
  et~al.}{2015}]{Querejeta2015}
{Querejeta} M.,  {Eliche-Moral} M.~C.,  {Tapia} T.,  {Borlaff} A.,
  {Rodr{\'\i}guez-P{\'e}rez} C.,  {Zamorano} J.,   {Gallego} J.,  2015, \mn@doi
  [\aap] {10.1051/0004-6361/201424303}, \href
  {https://ui.adsabs.harvard.edu/abs/2015A&A...573A..78Q} {573, A78}

\bibitem[\protect\citeauthoryear{{Roberts}, {Wardle}, {Lipnick}, {Selesnick}
  \& {Slutsky}}{{Roberts} et~al.}{2008}]{Roberts2008}
{Roberts} D.~H.,  {Wardle} J. F.~C.,  {Lipnick} S.~L.,  {Selesnick} P.~L.,
  {Slutsky} S.,  2008, \mn@doi [\apj] {10.1086/527544}, \href
  {https://ui.adsabs.harvard.edu/abs/2008ApJ...676..584R} {676, 584}

\bibitem[\protect\citeauthoryear{{Rossi}, {Bodo}, {Capetti}  \&
  {Massaglia}}{{Rossi} et~al.}{2017}]{Rossi2017}
{Rossi} P.,  {Bodo} G.,  {Capetti} A.,   {Massaglia} S.,  2017, \mn@doi [\aap]
  {10.1051/0004-6361/201730594}, \href
  {https://ui.adsabs.harvard.edu/abs/2017A&A...606A..57R} {606, A57}

\bibitem[\protect\citeauthoryear{{Rottmann}}{{Rottmann}}{2002}]{Rottmann2002}
{Rottmann} H.,  2002, PhD thesis, University of Bonn, Germany

\bibitem[\protect\citeauthoryear{{Rubinur}, {Das}, {Kharb}  \&
  {Honey}}{{Rubinur} et~al.}{2017}]{rubinur2017}
{Rubinur} K.,  {Das} M.,  {Kharb} P.,   {Honey} M.,  2017, \mn@doi [\mnras]
  {10.1093/mnras/stw2981}, \href
  {https://ui.adsabs.harvard.edu/abs/2017MNRAS.465.4772R} {465, 4772}

\bibitem[\protect\citeauthoryear{{Rubinur}, {Das}  \& {Kharb}}{{Rubinur}
  et~al.}{2019}]{rubinur2019}
{Rubinur} K.,  {Das} M.,   {Kharb} P.,  2019, \mn@doi [\mnras]
  {10.1093/mnras/stz334}, \href
  {https://ui.adsabs.harvard.edu/abs/2019MNRAS.484.4933R} {484, 4933}

\bibitem[\protect\citeauthoryear{{Rubinur}, {Kharb}, {Das}, {Rahna}, {Honey},
  {Paswan}, {Vaddi}  \& {Murthy}}{{Rubinur} et~al.}{2021}]{rubinur2021}
{Rubinur} K.,  {Kharb} P.,  {Das} M.,  {Rahna} P.~T.,  {Honey} M.,  {Paswan}
  A.,  {Vaddi} S.,   {Murthy} J.,  2021, \mn@doi [\mnras]
  {10.1093/mnras/staa3375}, \href
  {https://ui.adsabs.harvard.edu/abs/2021MNRAS.500.3908R} {500, 3908}

\bibitem[\protect\citeauthoryear{{Te{\c{s}}ileanu}, {Mignone}  \&
  {Massaglia}}{{Te{\c{s}}ileanu} et~al.}{2008}]{tesileanu2008}
{Te{\c{s}}ileanu} O.,  {Mignone} A.,   {Massaglia} S.,  2008, \mn@doi [\aap]
  {10.1051/0004-6361:200809461}, \href
  {https://ui.adsabs.harvard.edu/abs/2008A&A...488..429T} {488, 429}

\bibitem[\protect\citeauthoryear{Vaidya, Mignone, Bodo, Rossi  \&
  Massaglia}{Vaidya et~al.}{2018}]{Vaidya_2018}
Vaidya B.,  Mignone A.,  Bodo G.,  Rossi P.,   Massaglia S.,  2018, \mn@doi
  [The Astrophysical Journal] {10.3847/1538-4357/aadd17}, 865, 144

\bibitem[\protect\citeauthoryear{{Yates-Jones}, {Turner}, {Shabala}  \&
  {Krause}}{{Yates-Jones} et~al.}{2022}]{Yates2022}
{Yates-Jones} P.~M.,  {Turner} R.~J.,  {Shabala} S.~S.,   {Krause} M. G.~H.,
  2022, \mn@doi [\mnras] {10.1093/mnras/stac385}, \href
  {https://ui.adsabs.harvard.edu/abs/2022MNRAS.511.5225Y} {511, 5225}

\makeatother
\end{thebibliography}

\appendix
\section{Additional Simulation results} \label{sec:Additional Simulation results}
\subsection{Bigger domain run} \label{sec:Bigger domain}
In all of our runs highlighted in Table \ref{tab:simsetups}, we let the forward shock leave the computational domain that eventually weakens over time. Here, we showcase that this has not significantly affected the central structure, which is responsible for forming the S-shaped morphology. In this regard, we ran the simulation \textit{CaseB} with a bigger domain i.e. $-10 \leq x,z \leq 10$ and $-15 \leq y \leq 15$ so that at time 2.23 Myr the jet-cocoon structure remains well inside of the boundary. The resolution of the simulation is kept the same as \textit{CaseB} i.e. 15.6 pc to make comparisons between them. This requires the number of grids to enclose the domain as $256 \times 384 \times 256$ in the $x \times y \times z$ direction for the bigger domain run. 

\begin{figure}
    \centering
    \includegraphics[width=0.9\columnwidth]{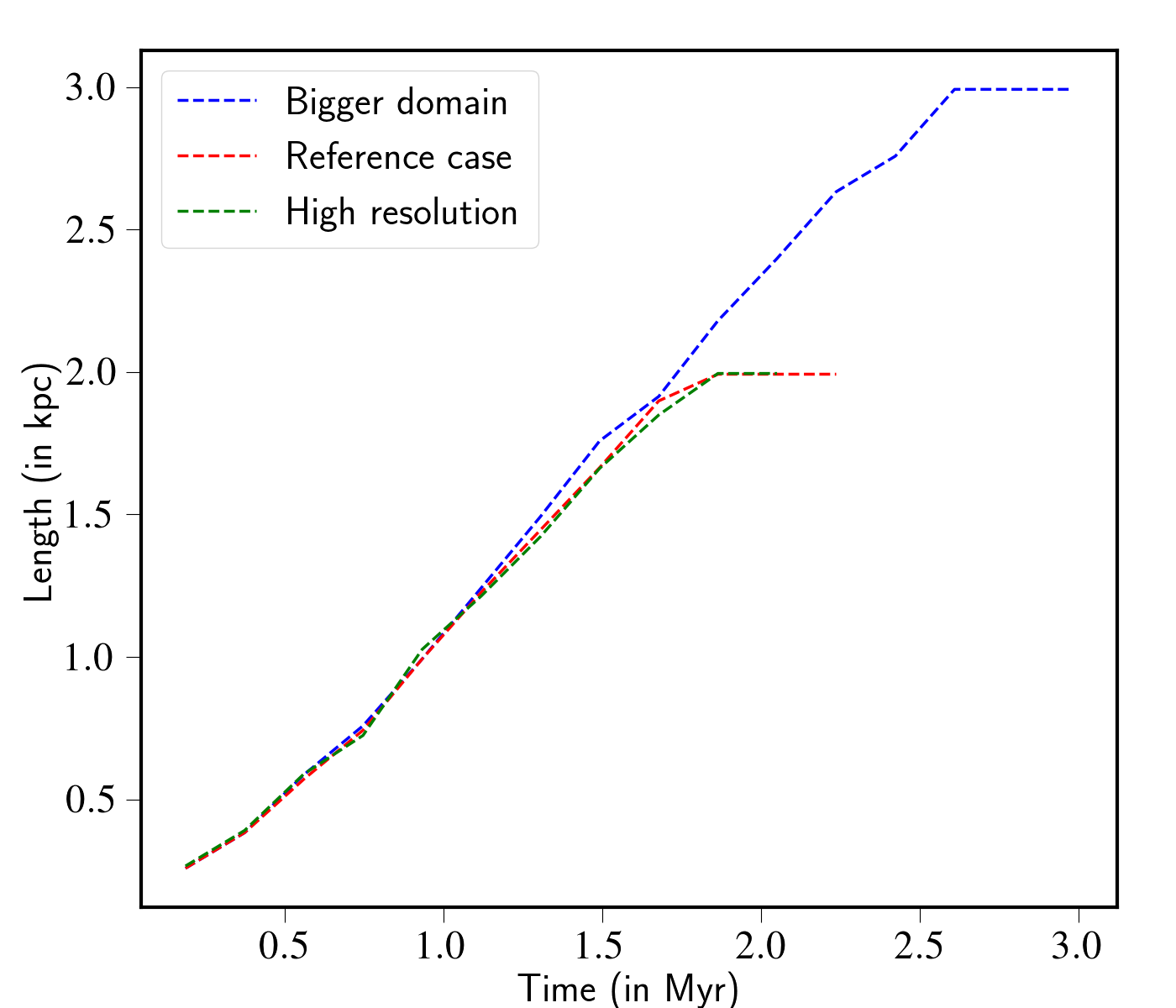}
    \includegraphics[width=0.9\columnwidth]{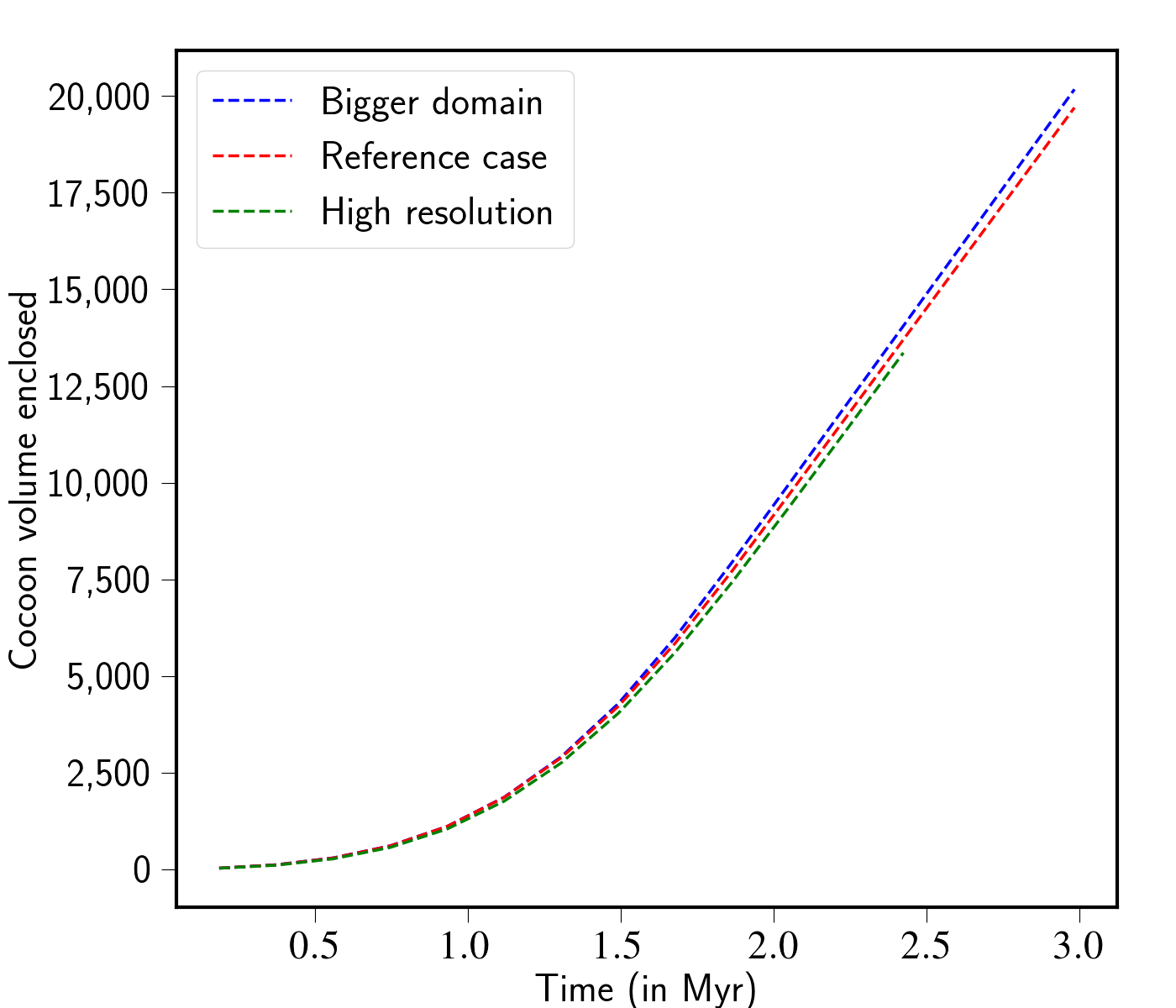}
    \caption{The above figures show the time variation of jet lengths (obtained by tracking the maximum pressure regions in the jet ambient interaction points) and cocoon volume (in computational unit) in the top and bottom, respectively. Here, three cases are considered for comparison: \textit{CaseB} of our simulation (reference case), same simulation but with bigger domain (Bigger domain run), and \textit{CaseB} with higher resolution (High resolution).}
    \label{fig:comparison}
\end{figure}

We compared both the size and shape of the cocoon obtained from these two runs by plotting jet length (tip point of the cocoon from the jet base) and cocoon volume with time, respectively. From Fig. \ref{fig:comparison} (top), we can see that the length of the jet as obtained from our reference run (\textit{CaseB}) follows the evolution shown by the curve obtained with the bigger domain run throughout the time. At time 1.9 Myr, the structure leaves the computational domain for \textit{CaseB} for which a flat line appears in the plot (red dashed line), however, the close match in the evolution of jet length (blue and red dashed lines) further showcase that our structure does not affect significantly even by letting a part of the cocoon leave the computational domain. To strengthen the evidence, we also have compared the volume enclosed within the cocoon between CaseB and bigger domain run. The volume is calculated in computational unit inside $-5 \leq x,z \leq 5$ and $-10 \leq y \leq 10$ (as this is the zone where we can make the comparison) which is shown in the bottom of Fig. \ref{fig:comparison}. We see also in this case that there is no systematic difference in the volume structures throughout the time for both cases.
\subsection{High resolution run}\label{sec:High Resolution}
In order to check that our results have reached a convergence in its properties in terms of resolution, we ran the simulation \textit{CaseB} with 1.5 times higher resolution. So, within the domain of $-5 \leq x,z \leq 5$ and $-10 \leq y \leq 10$ number of grid cells included is $192 \times 384 \times 192$ in the $x \times y \times z$ direction. This gives 39 grid cells within the jet diameter. In Fig. \ref{fig:comparison}, we show the comparison of the size and shape of the cocoon obtained from these two runs (i.e. \textit{CaseB} and higher resolution version of \textit{CaseB}) by plotting jet length and cocoon volume with time as discussed in Section \ref{sec:Bigger domain}, indicating a convergence in the results. This further indicates that our simulations highlighted in Table \ref{tab:simsetups} are performed with sufficient resolution.

\bsp	
\label{lastpage}
\end{document}